\documentclass{vldb}

\usepackage{amsmath, amssymb}
\usepackage{graphicx}
\usepackage{balance}
\usepackage{subfigure}
\usepackage{color}
\usepackage{mdwlist}
\usepackage{colortbl}
\usepackage[usenames,
             dvipsnames,
             svgnames,
             table]{xcolor}	% add support for color names

\usepackage{xspace}
\usepackage{float}
\usepackage{url}
\usepackage{textcomp}
\usepackage{booktabs}

\usepackage{hyperref}
\hypersetup{
	colorlinks=false,
	pdfborder={0 0 0},
}
\usepackage{breakurl}

\newtheorem{problem}{Problem}
\newtheorem{theorem}{Theorem}
\newtheorem{lemma}{Lemma}
\newtheorem{proposition}{Proposition}
\newtheorem{example}{Example}

\newcommand{\kw}[1]{{\ensuremath {\mathsf{#1}}}\xspace}

\newcommand{\eat}[1]{}

\newcommand{\bi}{\begin{itemize}}
\newcommand{\ei}{\end{itemize}}

\newcommand{\be}{\begin{enumerate}}
\newcommand{\ee}{\end{enumerate}}
\newcommand{\beqn}{\begin{eqnarray*}}
\newcommand{\eeqn}{\end{eqnarray*}}

\newcommand{\stitle}[1]{\vspace{1.5ex}\noindent{\bf #1}}

\newcommand{\ie}{\emph{i.e.,}\xspace}
\newcommand{\eg}{\emph{e.g.,}\xspace}

\newcounter{ccc}

\renewcommand{\texttt}[1]{{\small\textsf{#1}}}

\definecolor{gray}{rgb}{0.5,0.5,0.5}

\newcommand{\tgmine}{\kw{TGMiner}}
\newcommand{\gspan}{\kw{gSpan}}
\newcommand{\nodeset}{\kw{NodeSet}}
\newcommand{\freq}{\kw{freq}}
\newcommand{\ns}{\kw{nodeseq}}
\newcommand{\es}{\kw{edgeseq}}
\newcommand{\ens}{\kw{enhseq}}

\newcommand{\subprune}{\kw{SubPrune}}
\newcommand{\supprune}{\kw{SupPrune}}
\newcommand{\vfprune}{\kw{PruneVF2}}
\newcommand{\lscan}{\kw{LinearScan}}
\newcommand{\static}{\kw{Ntemp}}
\newcommand{\prunegi}{\kw{PruneGI}}
\newcommand{\interest}{\kw{interest}}
\newcommand{\sign}{\kw{sign}}
\newcommand{\syn}{\kw{SYN}}

\newcommand{\cut}[1]{}

\newcommand{\comment}[1]{}

\newcommand{\nop}[1]{}

\begin{document}
\pagenumbering{arabic}
\title{Behavior Query Discovery in System-Generated \\ Temporal Graphs}
\numberofauthors{1}
\author{\alignauthor Bo Zong$^{1}$ \hspace{4ex} Xusheng Xiao$^{1}$ \hspace{4ex} Zhichun Li$^{1}$ \hspace{4ex} Zhenyu Wu$^{1}$ \hspace{4ex} Zhiyun Qian$^{3}$ \hspace{4ex} Xifeng Yan$^{2}$ \hspace{4ex} Ambuj K. Singh$^{2}$ \hspace{4ex} Guofei Jiang$^{1}$ \\
\vspace{1ex}
\affaddr{$^{1}$NEC Labs America, Inc. \hspace{4ex} $^{2}$UC Santa Barbara  \hspace{4ex} $^{3}$UC Riverside} \\
\email{ \{\normalsize{bzong, xsxiao, zhichun, adamwu, gfj}\}@nec-labs.com \hspace{1ex} \{xyan, ambuj\}@cs.ucsb.edu \hspace{1ex} zhiyunq@cs.ucr.edu }
}

\maketitle

\begin{abstract}
Computer system monitoring generates huge amounts of logs that record the interaction of system entities. How to query such data to better understand system behaviors and identify potential system risks and malicious behaviors becomes a challenging task for system administrators due to the dynamics and heterogeneity of the data. System monitoring data are essentially heterogeneous temporal graphs with nodes being system entities and edges being their interactions over time.  Given the complexity of such graphs, it becomes time-consuming for system administrators to manually formulate useful queries in order to examine abnormal activities, attacks, and vulnerabilities in computer systems.

In this work, we investigate how to query temporal graphs and treat \emph{query formulation} as a discriminative temporal graph pattern mining problem. We introduce \tgmine to mine discriminative patterns from system logs, and these patterns can be taken as templates for building more complex queries. \tgmine leverages temporal information in graphs to prune graph patterns that share similar growth trend without compromising pattern quality.  Experimental results on real system data show that \tgmine is 6-32 times faster than baseline methods.  The discovered patterns were verified by system experts; they achieved high precision ($97\%$) and recall ($91\%$).
\end{abstract}

\section{Introduction}
\label{sec:intro}

\begin{figure}[thb]
\centering
\includegraphics[scale=0.5]{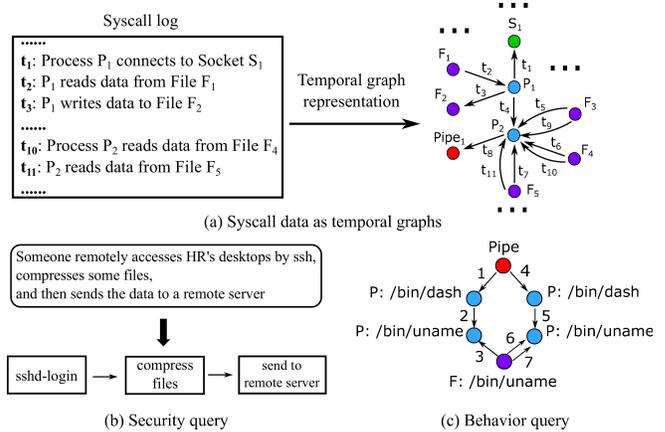}
\caption{Capture system activities: (a) syscall data represented as temporal graphs, (b) a security query, and (c) a discriminative subgraph pattern representing an sshd-login activity.}
\label{fig:intro-example}
\end{figure}

Computer systems are widely deployed to manage the business in industry and government. Ensuring the proper functioning of these systems is critical to the execution of the business. For example, if a system is compromised, the security of the customer data cannot be guaranteed; if certain components of a system have failures, the services hosted in the system may be interrupted. Maintaining the proper functioning of computer systems is a challenging task. System experts have limited visibility into systems, as the tools they use often give a partial view of the complex systems. This motivates the recent trend of leveraging system monitoring logs to offer intelligence in system management.

Temporal graphs are key data structures used to fuse information from heterogeneous sources~\cite{zheng2015detecting, zong2014kdd}. In computer systems, system monitoring data are represented as temporal graphs. For example, in cybersecurity, system call (syscall) logs provide a comprehensive way to capture system activities~\cite{backtracking2}.  Unlike its alternatives (\eg file access logs~\cite{splunk}, firewall~\cite{firewall}, and network monitoring~\cite{tcpwrapper}) which provide partial information and are application-specific, syscall logs cover all interactions among system entities (\eg processes, files, sockets, and pipes) over time. In Figure~\ref{fig:intro-example}(a), a syscall log contains a sequence of events each of which describes at which time what kind of interactions happened between which system entities. Note that this syscall log also forms an equivalent temporal graph.

While temporal graphs provide an intuitive way to visualize system behaviors, another important step is to query such data and leverage the results to better comprehend system status. The question is how! It is usually burdensome to formulate useful queries to search system-generated temporal graphs, as they are complex with many tedious low-level entities. Consider the following scenarios.

%\textbf{Cybersecurity}. 
\begin{example} [Cybersecurity]
To ensure the security of an enterprise system, a system expert wants to know if there exists any information stealthy activity in the human resource department over the weekend. A hypothetical activity could involve three steps: someone remotely accessed an HR desktop by ssh, compressed several files, and transferred them to a remote server. In order to find such activity, one can submit a query like Figure~\ref{fig:intro-example}(b) consisting of three components: ``sshd-login'', ``compress-files'', and ``send-to-remote-server'', and perform search over syscall logs like Figure~\ref{fig:intro-example}(a). However, such query cannot retrieve any useful information since the low-level entities (e.g., processes and files) recorded in the syscall logs cannot be directly mapped to any high-level activity like ``sshd-login'' or ``compress-files''. In order to locate all ``sshd-login'' activities, one has to know which processes or files are involved in ``sshd-login'' and in what order over time these low-level entities are involved in order to write a query. This becomes very time consuming. 

Besides querying risky behaviors, the formulated behavior queries can also be applied on the real-time monitoring data for surveillance and policy compliance checking.
\end{example}

%\textbf{Datacenter monitoring}. 
\begin{example} [Datacenter monitoring]
State-of-the-art system monitoring tools generate large-scale monitoring data as temporal graphs~\cite{zong2014kdd}, where nodes are system performance alerts, and edges indicate dependencies between alerts. While these alerts suggest low-level anomalies (\eg CPU usage is too high on server A, or there are too many full table joins on server B), system experts desire high-level knowledge about system behaviors: do these alerts result from \emph{disk failure} or \emph{abnormal database workload}? To search such high-level system behaviors, we have to know how alerts trigger each other and their temporal order~\cite{wang2010algorithmic}. It is a daunting task for system experts to manually formulate such queries.       
\end{example}

In addition, such query formulation problems also exist in other complex systems in the big data era.

%\textbf{Urban computing}. 
\begin{example} [Urban computing]
Modern cities generate a diverse array of data from heterogeneous sources, such as traffic flow, reports of sickness in cities, reports of food production, and so on~\cite{zheng2015detecting}. Information inferred from these sources are fused into temporal graphs, where nodes are events detected from different sources (\eg traffic jam, high sickness rate, and decrease in food production yield), edges indicate relationships between events (\eg two events are geographically close), and timestamps record when such relationships are detected. Domain experts are interested in high-level knowledge in urban systems: are these unusual events caused by river or air pollution? To formulate queries and search such knowledge, domain experts have to understand detailed temporal dependency patterns between events, which is extremely difficult for them.
\end{example}

Querying high-level system behaviors significantly reduces the complexity of evaluating system status, but it is quite difficult to formulate useful system behavior queries, referred to as \emph{behavior queries} in this paper, because of the big semantic gap between the high-level abnormal activities and the low-level footprints of such activities. To address this problem, one approach is to collect monitoring data of target behaviors (\eg ``sshd-login''), model the raw monitoring data by heterogeneous temporal graphs to , and use the full graphs to formulate queries. Unfortunately, the raw data can be large and noisy. To overcome this challenge, instead of using the full graphs, we identify the most discriminative patterns for target behaviors and treat them as queries. Such queries (\eg a few edges) are easier to interpret and modify, and are robust to noise. A discriminative pattern should frequently occur in the target activities, and rarely exist in other activities. One of the discriminative subgraph patterns for ``sshd-login'' is shown in Figure~\ref{fig:intro-example}(c), which includes a few nodes/edges and is more promising for querying ``sshd-login'' from syscall logs.

To this end, we formulate the behavior query construction problem as a \emph{discriminative temporal graph mining problem}:
Given a positive set and a negative set of temporal graphs, the goal is to find the temporal graph patterns with maximum discriminative score.
It is difficult to extend the existing mining techniques~\cite{jin2010gaia, yan2008mining} to solve this problem,
since they mainly focus on non-temporal graphs, not temporal graphs (detailed discussion in Section~\ref{sec:related:dpm}).

In this paper, we propose \tgmine that addresses the challenges to discriminative temporal graph pattern mining.

\begin{enumerate}

\item We have to consider both topology and edge temporal order while searching temporal graph space. To avoid redundant search, do we need another complex canonical labeling method like~\cite{jin2010gaia, yan2002gspan} for mining temporal graphs?  In our study, we find the temporal information in graphs allows us to explore temporal graph space in a more efficient manner. In particular, we propose a pattern growth algorithm without any complex canonical labeling. It guarantees that all promising patterns are covered, and no redundant search.

\item Since temporal graph space is huge, a naive exhaustive search is slow even for small temporal graphs. To speed up search, we first identify general cases where we can conduct pruning. Then we propose algorithms to minimize the overhead for discovering the pruning opportunities:
(1) By encoding temporal graphs into sequences, a light-weight algorithm based on subsequence tests is proposed to enable fast temporal subgraph tests; 
and (2) we compress residual graph sets into integers such that residual graph set equivalence tests are performed in constant time.

\end{enumerate}

Our major contributions are as follows. First, motivated by the need of queries in system monitoring applications, we identify a challenging query formulation problem in complex temporal graphs. Second, we propose the idea of using discriminative subgraph pattern mining to automatically formulate behavior queries, significantly easing the query formulation. Third, we develop \tgmine that leverages temporal information to enable fast pattern mining in temporal graphs. Experimental results on real data show that the behavior queries constructed by \tgmine are effective for behavior analysis in cybersecurity applications, with high precision $97\%$ and recall $91\%$, better than a non-temporal graph pattern based approach whose precision and recall are $83\%$ and $91\%$, respectively. For mining speed, \tgmine is 6-32 times faster than baseline methods.

%==========================Commented===========================
\eat{
\textbf{Network performance diagnosis}. Existing diagnostic tools for IPTV networks produce temporal graphs where nodes are network events and edges represent event dependencies~\cite{mahimkar2009towards}. Although the raw events reveal how small components in the network changed their states, service providers expect to understand whether the observed events suggest any high-level activities (\eg slow digital video recording or failures in parental control). Temporal dependency patterns between events provide possible queries to search high-level events, but it is difficult for domain experts to manually identify such patterns.
}
\section{Problem Formulation}
\label{sec:problem-statement}

\begin{figure*}[tbh]
\center
\includegraphics[scale=0.55]{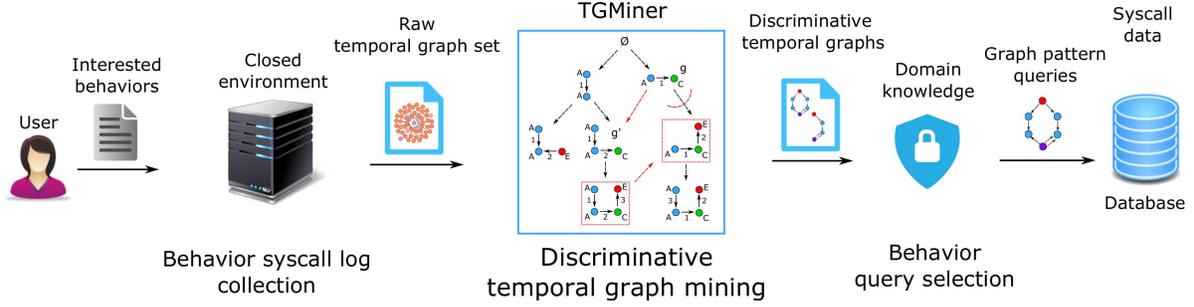}
\caption{Behavior Query Formulation Pipeline}
\label{fig:design-overview}
\end{figure*}

As described earlier, our goal is to mine skeletons for behavior queries from temporal graphs derived from system monitoring data. In the following, we focus on cybersecurity applications for the ease of presentation, but the proposed ideas and techniques can also be applied to applications in other domains. 

\textbf{Temporal graph}.
A temporal graph $G$ is represented by a tuple $(V, E, A, T)$, where (1) $V$ is a node set;
(2) $E \subset V \times V \times T$ is a set of directed edges that are totally ordered by their timestamps;
(3) $A: V \rightarrow \Sigma$ is a function that assigns labels to nodes ($\Sigma$ is a set of node labels);
and (4) $T$ is a set of possible timestamps, non-negative integers on edges.

In practice, the syscall data of a behavior instance is collected from a controlled environment, where only one target behavior is performed. In most cases, the syscall data of a target behavior forms a temporal graph of no more than a few thousand of nodes/edges.

We target temporal graphs with total edge order in this work. Our empirical results show that this model performs quite well in identifying basic security behaviors, as these behaviors are usually finished by one thread. It also possesses computation advantages, in comparison with more complex temporal graph models. For the cases of concurrent edges, we provide a discussion in Section~\ref{sec:concurrent}.

\begin{figure}[thb]
\center
\includegraphics[scale=0.5]{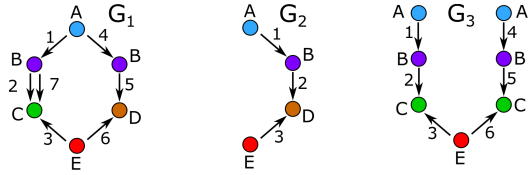}
\caption{Temporal graphs: (1) $G_2$ is a temporal subgraph of $G_1$ ($G_2 \subseteq_t G_1$); and (2) $G_1$ and $G_2$ are T-connected, while $G_3$ is non T-connected.}
\label{fig:pd-example}
\end{figure}

Figure~\ref{fig:pd-example} shows three temporal graphs. Note that multi-edges are allowed in temporal graphs as shown in $G_1$. For simplicity, we examine temporal graphs with only node labels and edge timestamps in this paper. Our algorithms also work for temporal graphs with edge labels.

\textbf{Temporal graph pattern}. A temporal graph pattern $g = (V, E, A, T)$ is a temporal graph, where $\forall t \in T, 1 \leq t \leq |E|$.

Unlike general temporal graphs where timestamps could be arbitrary non-negative integers, timestamps in temporal graph patterns are aligned (from 1 to $|E|$) and only total edge order is kept. In the following discussion, we use upper case letters (such as $G$) to represent temporal graph data, and use lower case letters (such as $g$) to represent abstract temporal graph patterns.

As it is ineffective and inefficient to take the entire raw graph as a behavior query, it will be better to use its discriminative subgraphs to capture the footprint of a behavior.

\textbf{Temporal subgraph}.
Given two temporal graphs $G = (V, E, A, T)$ and $G' = (V', E', A', T')$, $G \subseteq_t G'$ if and only if there exist two \emph{injective} functions $f: V \rightarrow V'$ and $\tau: T \rightarrow T'$ such that
(1) \emph{node mapping}: $\forall u \in V$, $A(u) = A'(f(u))$;
(2) \emph{edge mapping}: $\forall (u, v, t) \in E$, $\big( f(u), f(v), \tau(t) \big) \in E'$;
and (3) \emph{edge order preserved}: $\forall (u_1, v_1, t_1), (u_2, v_2, t_2) \in E$, $\sign(t_1 - t_2) = \sign(\tau(t_1) - \tau(t_2))$.

$G'$ is a \emph{match} of $G$ denoted as $G' =_t G$, when $f$ and $\tau$ are \emph{bijective} functions.

Figure~\ref{fig:pd-example} shows an example of a temporal subgraph where $G_2 \subseteq_t G_1$. In particular, the subgraph in $G_1$ formed by edges of timestamps $4$, $5$, and $6$ is a match of $G_2$.

\textbf{Pattern frequency}.
Given a set of temporal graphs $\mathbb{G}$ and a temporal graph pattern $g$, the frequency of $g$ with respect to $\mathbb{G}$ is defined as
\[
\freq(\mathbb{G}, g) = \frac{ |\{ G \mid g \subseteq_t G \wedge G \in \mathbb{G} \} | }{|\mathbb{G}|}.
\]

Moreover, we differentiate two types of connected graphs among temporal graphs.

\textbf{T-connected temporal graph}.
A temporal graph $G = (V, E, A, T)$ is T-connected if $\forall (u, v, t) \in E$, the edges whose timestamps are smaller than $t$ form a connected graph.

In Figure~\ref{fig:pd-example}, $G_1$ and $G_2$ are T-connected temporal graphs while $G_3$ is not, because the graph formed by edges with timestamps smaller than $5$ is disconnected.

In this work, we focus on mining T-connected temporal graph patterns for the following reasons. First, in pattern growth, T-connected patterns remain connected, while non T-connected patterns might be disconnected during the growth process, resulting in formidable explosion of pattern search space. Second, any non T-connected temporal graph is formed by a set of T-connected temporal graphs. In practice, we can use a single T-connected pattern or a set of T-connected patterns that in all could be a non T-connected pattern to form a behavior query.
In the rest of this paper, T-connected temporal graphs are referred to as connected temporal graphs without ambiguity.
Next we define the discriminative temporal graph pattern mining problem.

\begin{problem}
Given a set of positive temporal graphs $\mathbb{G}_p$ and a set of negative temporal graphs $\mathbb{G}_n$, the goal is to find the connected temporal graph patterns $g^{*}$ with maximum $F\big( \freq(\mathbb{G}_p, g^{*}), \freq(\mathbb{G}_n, g^{*})  \big)$, where $F(x, y)$ is a discriminative score function with partial (anti-)monotonicity: (1) when $x$ is fixed, $y$ is smaller, $F(x, y)$ is larger;
and (2) when $y$ is fixed, $x$ is larger, $F(x, y)$ is larger.
\label{def:problem}
\end{problem}

$F(x, y)$ covers many widely used score functions including G-test, information gain, and so on~\cite{yan2008mining}. In practice, one could pick a discriminative score function that satisfies partial (anti-)monotonicity and best fits his/her query formulation task. Note that for the ease of presentation, the discriminative score of a pattern $g$ is also denoted as $F(g)$ in the following discussion.

\textbf{Behavior query formulation pipeline}.
Figure~\ref{fig:design-overview} shows a pipeline of collecting syscall logs for behaviors, finding patterns, and using them to construct graph pattern queries for searching behaviors from syscall data (temporal graphs) and retrieving interesting security knowledge. We take the behavior of sshd-login as an example.

The first step is to form input data. Relatively clean syscall logs for sshd-login are crawled from a closed environment, where sshd-login is independently run multiple times. Additionally, we also collect syscall logs where sshd-login is not performed and treat them as background syscall logs.  The input of mining sshd-login behavior patterns is formed as follows: (1) the raw syscall logs of sshd-login are treated as a set of positive temporal graphs $\mathbb{G}_p$ and (2) the raw background syscall logs are treated as a set of negative temporal graphs $\mathbb{G}_n$. In practice, we can also use the syscall logs for normal or abnormal sshd-login (\eg intrusion) as positive datasets, which will generate graph pattern queries for normal and abnormal behaviors.

Given $\mathbb{G}_p$ and $\mathbb{G}_n$, \tgmine finds the most discriminative temporal graph patterns for sshd-login. To identify the patterns that best serve the purpose of behavior search, the patterns discovered by \tgmine are further ranked by domain knowledge, including semantic/security implication on node labels and node label popularity among monitoring data. Top ranked patterns are then selected as queries to search sshd-login activities from a repository of syscall log data and see if there are abnormal/suspicious activities, \eg too many times of sshd-login over a Saturday night.

\textbf{\tgmine overview}. Our mining algorithm includes two key components: pattern growth and pattern space pruning.

Pattern growth guides the search in pattern space. It conducts depth-first search: starting with an empty pattern, growing it into a one-edge pattern, and exploring all possible patterns in its branch. When one branch is completely searched, the algorithm continues to explore the branches initiated by other one-edge patterns. The key challenges in this component is how to avoid repeated pattern search and how to cover the whole pattern space.

Pattern space pruning is a crucial step to speed mining processes. The underlying pattern space could be large, and a naive search algorithm cannot scale. Therefore, effective pruning algorithms are desired. The key problems in this component is how to identify general pruning opportunities and how to minimize the overhead in pruning.

%In Section~\ref{sec:growth} and Section~\ref{sec:pruning}, we discuss how we address the challenges in these two components in details.

%=========================removed contents================================

\section{Temporal Graph Growth}
\label{sec:growth}

In this section,
we discuss the pattern growth algorithm in {\tgmine}.
In particular,
we demonstrate (1) how temporal information in graphs enables efficient pattern growth without repetition,
and (2) the general principles to grow temporal graph patterns so that all possible connected temporal graph patterns will be covered.

\subsection{Growth without Repetition}
\label{sec:growth:forward}

Pattern growth is more efficient in temporal graphs, compared with its counterpart in non-temporal graphs.

It is costly to conduct pattern growth for non-temporal graphs.
To grow a non-temporal pattern to a specific larger one,
there exist a combinatorial number of ways.
In order to avoid repeated computation, we need extra computation to confirm whether one pattern is a new pattern or is a discovered one.
This results in high computation cost, as graph isomorphism is inevitably involved.
To reduce the overhead, various canonical labeling techniques along with their sophisticated pattern growth algorithms~\cite{huan2003efficient, jin2010gaia, yan2002gspan} have been proposed, but the cost is still very high because of the intrinsic complexity in graph isomorphism.

Unlike non-temporal graphs, 
we can develop efficient algorithms for temporal graph pattern growth.
First, we can decide whether two temporal graph patterns are identical in linear time. 
Second, these is at most one possible way to grow a temporal graph pattern to a specific larger one.
Next, we show why these properties stand.  

The computation advantages of temporal graphs originate from the following property.
\begin{lemma}
Let $g_1$ and $g_2$ be temporal graph patterns. If $g_1 =_t g_2$, the mappings $f$ and $\tau$ between them are unique.
\label{lm:unique}
\end{lemma}
\begin{proof}
The proof is detailed in Appendix A.
\end{proof}

With Lemma~\ref{lm:unique}, we further prove the efficiency of determining whether two temporal graph patterns are matched.

\begin{lemma}
If $g_1$ and $g_2$ are temporal graph patterns,
then $g_1 =_t g_2$ can be determined in linear time.
\label{lm:linear-match}
\end{lemma}
\begin{proof}
The proof is detailed in Appendix B.
\end{proof}

Pattern growth for temporal graphs will be more efficient, when it is guided by \emph{consecutive growth}.

\begin{figure}[thb]
\center
\includegraphics[scale=0.5]{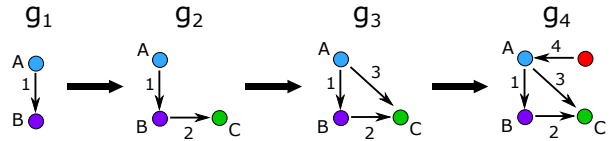}
\caption{An example of consecutive growth}
\label{fig:growth-forward}
\end{figure}

\textbf{Consecutive growth}.
Given a connected temporal graph pattern $g$ of edge set $E$ and an edge $e' = (u', v', t')$,
adding $e'$ into $g$ is consecutive growth,
if (1) it results in another connected temporal graph pattern;
and (2) $t' = |E|+1$.

Figure~\ref{fig:growth-forward} demonstrates how $g_1$ grows to $g_4$ by consecutive growth. 
Consecutive growth guarantees a connected temporal graph pattern will form another connected temporal graph pattern without repetition.

\begin{lemma}
Let $g_1$ and $g_2$ be connected temporal graph patterns with $g_1 \subseteq g_2$.
If pattern growth is guided by consecutive growth,
then (1) either there exists a unique way to grow $g_1$ into $g_2$,
(2) or there is no way to grow $g_1$ into $g_2$.
\label{lm:forward}
\end{lemma}

\begin{proof}
The proof is detailed in Appendix C.
\label{proof:lm:forward}
\end{proof}

Unlike mining non-temporal graphs, 
by Lemma~\ref{lm:linear-match} and~\ref{lm:forward},
we can avoid repeated pattern search without using any sophisticated canonical labeling or complex pattern growth algorithms~\cite{jin2010gaia, yan2008mining}.
Next, we show the three growth options that one needs to cover the whole pattern space.

\subsection{Growth Options}
\label{sec:growth:option}

\begin{figure}[thb]
\center
\includegraphics[scale=0.5]{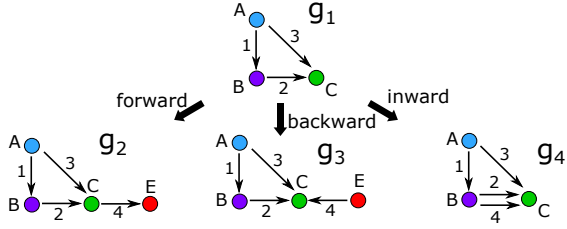}
\caption{Three growth options}
\label{fig:growth-option}
\end{figure}

To guarantee the quality of discovered patterns, we need to ensure our search algorithm can cover the whole pattern space.
In this work, we identify three growth options to achieve the completeness.
Let $g$ be a connected temporal graph pattern with node set $V$.
We can grow $g$ by consecutive growth as follows.
\begin{itemize}
\item Forward growth: 
growing $g$ by an edge $(u, v, t)$ is a forward growth
if $u \in V$ and $v \notin V$.

\item Backward growth:
growing $g$ by an edge $(u, v, t)$ is a backward growth
if $u \notin V$ and $v \in V$.

\item Inward growth:
growing $g$ by an edge $(u, v, t)$ is an inward growth
if $u \in V$ and $v \in V$.
\end{itemize}
Figure~\ref{fig:growth-option} illustrates the three growth options.
Note that inward growth allows multi-edges between node pairs.

The three growth options provide a guidance to conduct a complete search over pattern space.

\begin{theorem}
Let $\mathcal{A}$ be a search algorithm following consecutive growth with forward, backward, and inward growth.
Algorithm $\mathcal{A}$ guarantees (1) a complete search over pattern space,
and (2) no pattern will be searched more than once.
\label{thm:growth}
\end{theorem}

\begin{proof}
The proof is detailed in Appendix D.
\label{proof:theorem1}
\end{proof}

Theorem~\ref{thm:growth} provides a naive exhaustive approach to mining discriminative temporal graph patterns. 
However, the underlying pattern space is usually huge,
and the naive method suffers from poor mining speed.
To speed up the mining process, we propose pruning algorithms in Section~\ref{sec:pruning}.

%====================================Removed contents=====================================

\section{Pruning Temporal Graph Space}
\label{sec:pruning}

In this section, we investigate how to prune search space by the unique properties in temporal graphs.
First, we explore the general cases where we can prune unpromising branches.  %(Section~\ref{sec:pruning:cases}).
Next, we identify the major overhead for discovering pruning opportunities, and leverage the temporal information in graphs to minimize the overhead.

\subsection{Naive Pruning Conditions}
\label{sec:pruning:naive}

A straightforward pruning condition is to consider the upper bound of a pattern's discriminative score.
Given a temporal graph pattern $g$, the upper bound of $g$ indicates the largest possible discriminative score that could be achieved by $g$'s supergraphs.
Let $\mathbb{G}_p$ and $\mathbb{G}_n$ be positive graph set and negative graph set, respectively.
Since $\forall g \subseteq_t g'$, $\freq(\mathbb{G}_p, g') \leq \freq(\mathbb{G}_p, g)$ and $\freq(\mathbb{G}_n, g') \geq 0$,
we can derive the following upper bound,
\[
F\big( \freq(\mathbb{G}_p, g'), \freq(\mathbb{G}_n, g')  \big) \leq F\big( \freq(\mathbb{G}_p, g), 0 \big).
\]
This upper bound is theoretically tight; however, it is ineffective for pruning in practice~\cite{yan2008mining}.
In the rest of this section, we discuss general pruning opportunities inspired by temporal sub-relations. 

\subsection{Pruning by Temporal Sub-relations}
\label{sec:pruning:cases}

\begin{figure}[thb]
\center
\includegraphics[scale=0.45]{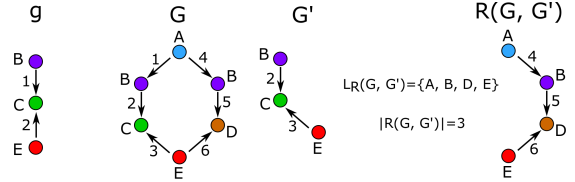}
\caption{An example of residual graph}
\label{fig:pruning-residual}
\end{figure}

For a temporal graph pattern, we denote the graph data that need to be considered for growing this pattern as a set of \emph{residual graphs}.

Let $G'$ be a subgraph of $G$.
If we remove the edges in $G$ whose timestamps are no larger than the largest edge timestamp in $G'$,
then we form a residual graph.

\textbf{Residual graph}.
Given a temporal graph $G = (V, E, A, T)$ and its subgraph $G' = (V', E', A', T')$,
$R(G, G') = (V_R, E_R, A_R, T_R)$ is $G$'s residual graph with respect to $G'$,
where (1) $E_R \subset E$ satisfies $\forall (u_1, v_1, t_1) \in E_R, (u_2, v_2, t_2) \in E'$, $t_1 > t_2$;
and (2) $V_R$ is the set of nodes that are associated with edges in $E_R$.
The size of $R(G, G')$ is defined as $| R(G, G') | = | E_R |$ (\ie the number of edges in $R(G, G')$).

Given a residual graph $R(G, G')$, its residual node label set is defined as $L_R(G, G') = \{A_R(u) \mid \forall u \in V_R \}$.

Figure~\ref{fig:pruning-residual} demonstrates an example of residual graph,
where $G'$ is a subgraph of $G$, $R(G, G')$ is $G$'s residual graph with respect to $G'$, and $L_R(G, G')$ is its residual node set.

Let $\mathbb{M}(G, g)$ be a set including all the subgraphs in $G$ that match a temporal graph pattern $g$.
Given $\mathbb{G}_p$ and $g$, we define positive residual graph set $\mathbb{R}(\mathbb{G}_p, g)$ as
\[ 
\mathbb{R}(\mathbb{G}_p, g) = \bigcup_{G \in \mathbb{G}_p} \{ R(G, G') \mid G' \in \mathbb{M}(G, g) \}.
\]

Given $\mathbb{R}(\mathbb{G}_p, g)$,  its residual node label set $\mathbb{L}(\mathbb{G}_p, g)$ is defined as
\[
\mathbb{L}(\mathbb{G}_p, g) = \bigcup_{G \in \mathbb{G}_p} \bigcup_{G' \in \mathbb{M}(G, g)} L_R(G, G').
\] 

Similarly, we can define negative residual graph set $\mathbb{R}(\mathbb{G}_n, g)$ and its residual node label set $\mathbb{L}(\mathbb{G}_n, g)$.

Next, we introduce another important property that helps us identify pruning opportunities.

\begin{proposition}
Given a temporal graph set $\mathbb{G}$ and two temporal graph patterns $g_1 \subseteq_t g_2$,
if $\mathbb{R}(\mathbb{G}, g_1) = \mathbb{R}(\mathbb{G}, g_2)$, then the node mapping between $g_1$ and $g_2$ is unique. 
\label{prop:proposition}
\end{proposition}
\begin{proof}
We sketch the proof as follows. First, we prove if the node mapping between $g_1$ and $g_2$ is not unique, any match for $g_2$ includes multiple matches for $g_1$. Then we prove if any match for $g_2$ includes multiple matches for $g_1$, $\mathbb{R}(\mathbb{G}, g_1) = \mathbb{R}(\mathbb{G}, g_2)$ will never be true.
The detailed proof is stated in Appendix E.
\end{proof}

In the following, we present \emph{subgraph pruning} and \emph{supergraph pruning}.
In the following discussion, for a temporal graph pattern $g$,
we use $g$'s branch to refer to the space of patterns that are grown from $g$,
and use $F^*$ to denote the largest discriminative score discovered so far. 

Let $g_1$ and $g_2$ be temporal graph patterns where $g_1$ is discovered before $g_2$.
If (1) $g_2$ is a temporal subgraph of $g_1$; 
(2) they share identical positive residual graph sets;
and (3) for those nodes in $g_1$ that cannot match to any nodes in $g_2$, their labels never appear in $g_2$'s residual node label set,
then we can conduct subgraph pruning on $g_2$.

\textbf{Subgraph pruning}.
Given a discovered pattern $g_1= (V_1, E_1, A_1, T_1)$ and a pattern $g_2$ of node set $V_2$,
if (1) $g_2 \subseteq_t g_1$,
(2) $\mathbb{R}(\mathbb{G}_p, g_2) = \mathbb{R}(\mathbb{G}_p, g_1)$,
and (3) $\mathbb{L}(\mathbb{G}_p, g_2) \cap L_{g_1 \setminus g_2} = \emptyset$ (where $L_{g_1 \setminus g_2} = \{ A_1(u) \mid \forall u \in V_1 \setminus V'_1 \}$ and $V'_1 \subseteq V_1$ is the set of nodes that map to nodes in $V_2$),
we can prune the search on $g_2$'s branch, if the largest discriminative score for patterns in $g_1$'s branch is smaller than $F^*$.

\begin{figure}[thb]
\center
\includegraphics[scale=0.45]{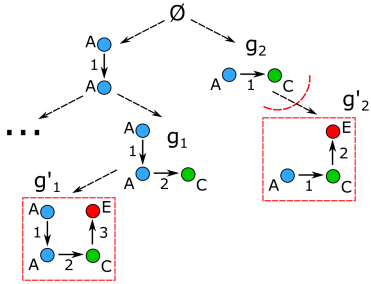}
\caption{Subgraph pruning}
\label{fig:pruning-subgraph}
\end{figure}

Figure~\ref{fig:pruning-subgraph} illustrates the idea in subgraph pruning. 
In the mining process,
we reach a pattern $g_2$,
and we also notice that there exists a discovered pattern $g_1$,
which satisfies the conditions in subgraph pruning. 
Therefore,
pattern growth in $g_1$'s branch suggests how to grow $g_2$ to larger patterns (\eg growing $g_1$ to $g'_1$ indicates we can grow $g_2$ to $g'_2$).
Since none of the patterns in $g_1$'s branch have the score $F^*$,
the patterns in $g_2$'s branch cannot be the most discriminative ones as well, which can be safely pruned.

\begin{lemma}
Subgraph pruning prunes pattern space without missing any of the most discriminative patterns.   
\label{lm:subgraph}
\end{lemma}

\begin{proof}
The proof is detailed in Appendix F.
\label{proof:subgraph}
\end{proof}

Similar to subgraph pruning, we can perform supergraph pruning. 
Let $g_1$ and $g_2$ be temporal graph patterns where $g_1$ is discovered before $g_2$.
If (1) $g_1$ is a temporal subgraph of $g_2$,  (2) they share identical positive and negative residual graph sets, and (3) they have the same number of nodes, we can conduct supergraph pruning on $g_2$.

\textbf{Supergraph pruning}.
Given two patterns $g_1$ and $g_2$, 
where $g_1$ is discovered before $g_2$ and $g_2$ is not grown from $g_1$,
if (1) $g_2 \supseteq_t g_1$, 
(2) $\mathbb{R}(\mathbb{G}_p, g_2) = \mathbb{R}(\mathbb{G}_p, g_1)$,
(3) $\mathbb{R}(\mathbb{G}_n, g_2) = \mathbb{R}(\mathbb{G}_n, g_1)$,
and (4) $g_2$ and $g_1$ have the same number of nodes,
the search in $g_2$'s branch can be safely pruned,
if the largest discriminative score for $g_1$'s branch is smaller than $F^*$. 

\begin{figure}[thb]
\center
\includegraphics[scale=0.45]{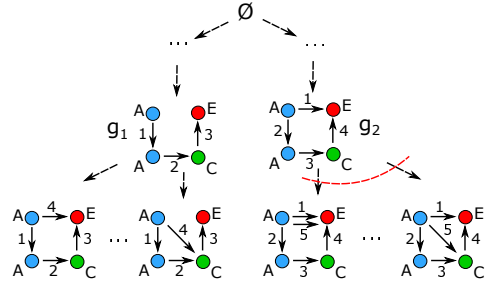}
\caption{Supergraph pruning}
\label{fig:pruning-supergraph}
\end{figure}

Figure~\ref{fig:pruning-supergraph} demonstrates the idea in supergraph pruning.
In the mining process,
a temporal graph pattern $g_2$ is reached,
and there is another pattern $g_1$ discovered before $g_2$, which satisfies the conditions in supergraph pruning.
Therefore, the growth knowledge in $g_1$'s branch suggests how to grow $g_2$ to larger patterns.
Since none of the patterns in $g_1$'s branch are the most discriminative,
we can infer the patterns in $g_2$'s branch are unpromising as well,
and the search in $g_2$'s branch can be safely pruned.

\begin{proposition}
Supergraph pruning prunes pattern space without missing the most discriminative patterns. 
\label{prop:supergraph}
\end{proposition}

\begin{proof}
The proof is detailed in Appendix G.
\label{proof:supergraph}
\end{proof}

Lemma~\ref{lm:subgraph} and Proposition~\ref{prop:supergraph} lead to the following theory.

\begin{theorem}
Performing subgraph pruning and supergraph pruning guarantees the most discriminative patterns will still be preserved.
\label{thm:pruning}
\end{theorem}

Theorem~\ref{thm:pruning} identifies general cases where we can conduct pruning in temporal graph space; 
however, it works only if the overhead for discovering these pruning opportunities is small enough.
The major overhead of subgraph pruning and supergraph pruning comes from two sources:
(1) temporal subgraph tests (\eg $g_2 \subseteq_t g_1$),
and (2) residual graph set equivalence tests (\eg $\mathbb{R}(\mathbb{G}_p, g_2) = \mathbb{R}(\mathbb{G}_p, g_1)$).
For example, to mine patterns for ``sshd-login'' behavior, the mining process involves more than 70M temporal subgraph tests and 400M residual graph set equivalence tests. 
These overhead is not negligible, and it significantly degrades the efficiency of the pruning algorithm.
Next, we investigate how to minimize these overhead.

\subsection{Temporal Subgraph Test}
\label{sec:pruning:sub-test}
In this section, we discuss how to leverage the temporal information in graphs to minimize the overhead from temporal subgraph tests.
First, we propose an encoding scheme that represent temporal graphs by sequences. Second, we develop a light-weight algorithm based on subsequence tests.

Similar to subgraph test for non-temporal graphs, it is difficult to efficiently perform temporal subgraph tests.

\begin{proposition}
Given two temporal graphs $g$ and $g'$, it is NP-complete to decide $g \subseteq_t g'$.
\label{prop:hardness}
\end{proposition}

\begin{proof}
We sketch the proof as follows. First, we prove the NP-hardness by reducing clique problem~\cite{michael1979computers} to temporal subgraph test problem. By transforming non-temporal graphs into temporal graphs in polynomial time, we show we solve clique problem by temporal subgraph tests, which implies temporal subgraph test problem is at least as hard as clique problem. Second, we prove temporal subgraph test problem is NP by showing we can verify its solution in polynomial time.
The proof is detailed in Appendix H.
\end{proof}

While existing algorithms~\cite{cordella2004sub, sun2012efficient, zhang2010sapper} for non-temporal subgraphs provide possible solutions,
temporal information in graphs suggests the existence of a faster solution.
\begin{enumerate}
\item Since edges are totally ordered in temporal graphs, it is possible to encode temporal graphs into sequences.

\item After temporal graphs are represented as sequences, it is possible to enable faster temporal subgraph tests using efficient subsequence tests.
\end{enumerate}
Based on these insights, we propose a light-weight temporal subgraph test algorithm.
In particular, this algorithm consists of two components:
(1) a sequence-based temporal graph representation,
and (2) a temporal subgraph test algorithm based on subsequence tests.

Before we dive into the technical details, we review the definition of subsequence.
Let $s_1 = (a_1, a_2, ..., a_n)$ and $s_2 = (b_1, b_2, ..., b_m)$ be two sequences.
If there exist $1 \leq i_1 < i_2 < ... < i_n \leq m$ such that $\forall 1 \leq j \leq n$, $a_j = b_{i_j}$, $s_1$ is a subsequence of $s_2$, denoted as $s_1 \sqsubseteq s_2$.

\textbf{Sequence-based representation}.
A temporal graph pattern $g$ can be represented by two sequences.
\begin{itemize}
\item Node sequence $\ns(g)$ is a sequence of labeled nodes. 
Given $g$ is traversed by its edge temporal order,
nodes in $\ns(g)$ are ordered by their first visited time.
Any node of $g$ appears only once in $\ns(g)$.

\item Edge sequence $\es(g)$ is a sequence of edges in $g$, where edges are ordered by their timestamps;
\end{itemize}

\begin{figure}[thb]
\center
\includegraphics[scale=0.42]{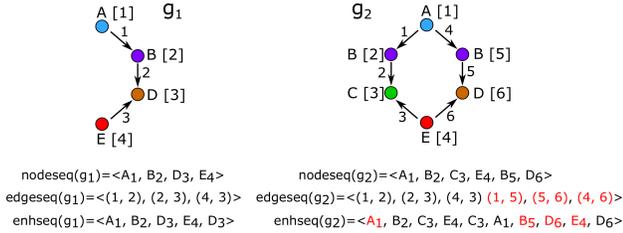}
\caption{Sequence-based temporal graph representation
and temporal subgraph tests}
\label{fig:pruning-sequence-test}
\end{figure}

Figure~\ref{fig:pruning-sequence-test} illustrates examples of sequence-based representation.
In $g_1$ and $g_2$, node labels are represented by letters, and nodes of the same labels are differentiated by their node IDs represented by integers in brackets.
Node labels in $\ns$ are associated with node IDs as subscripts.
Note that when we compare node labels, their subscripts will be ignored (\ie $\forall i, j, B_i = B_j$).
Each edge in $\es$ is represented by the following format $(id(u), id(v))$, where $id(u)$ is the source node ID and $id(v)$ is the destination node ID.

Given two temporal graphs $g_1$ and $g_2$, if $g_1 \subseteq_t g_2$, we expect $\ns(g_1) \sqsubseteq \ns(g_2)$ and $\es(g_1) \sqsubseteq \es(g_2)$. However, when $g_1 \subseteq_t g_2$, $\ns(g_1) \sqsubseteq \ns(g_2)$ may not be true. As shown in Figure~\ref{fig:pruning-sequence-test}, $\ns(g_1) \not \sqsubseteq \ns(g_2)$ because the first visited time of the node with label $E$ is inconsistent in $g_1$ and $g_2$.

To address this issue, we propose enhanced node sequence $\ens$.
Let $g$ be a temporal graph.
$\ens(g)$ is a sequence of labeled nodes in $g$.
Given $g$ is traversed by its edge temporal order, $\ens(g)$ is constructed by processing each edge $(u, v, t)$ as follows.
(1) If $u$ is the last added node in the current $\ens(g)$, or $u$ is the source node of the last processed edge, $u$ will be skipped; otherwise, $u$ will be added into $\ens(g)$.
(2) Node $v$ will be always added into $\ens(g)$.
Note that nodes in $g$ might appear multiple times in $\ens(g)$.
Figure~\ref{fig:pruning-sequence-test} shows the enhanced node sequences of $g_1$ and $g_2$.

With the support from enhanced node sequence,
we are ready to build the connection between temporal subgraph tests and subsequence tests.

\begin{lemma}
Two temporal graphs $g_1 \subseteq_t g_2$ if and only if
\begin{enumerate}
\item $\ns(g_1) \sqsubseteq \ens(g_2)$, 
where the underlying match forms an injective node mapping $f_{s}$ from nodes in $g_1$ to nodes in $g_2$;

\item  $f_{s}(\es(g_1)) \sqsubseteq \es(g_2)$,
where $f_{s}(\es(g_1))$ is an edge sequence where the nodes in $g_1$ are replaced by the nodes in $g_2$ via the node mapping $f_{s}$.  
\end{enumerate}
\label{lm:subseq}
\end{lemma}

\begin{proof}
The proof is detailed in Appendix I.
\label{proof:subseq}
\end{proof}

As shown in Figure~\ref{fig:pruning-sequence-test}, 
$g_1$ and $g_2$ are two temporal graphs satisfying $g_1 \subseteq_t g_2$.
The node sequence of $g_1$ is a subsequence of the enhanced node sequence of $g_2$ with the injective node mapping 
$f_{s}(1) = 1$, $f_{s}(2) = 5$, $f_{s}(3) = 6$, and $f_{s}(4) = 4$.
Therefore, we obtain $f_{s}(\es(g_1)) = \langle (1, 5), (5, 6), (4, 6) \rangle$ so that $f_{s}(\es(g_1)) \sqsubseteq \es(g_2)$.

\textbf{Subsequence-test based algorithm}.
A temporal subgraph test algorithm is derived from Lemma~\ref{lm:subseq}. Given temporal graphs $g_1$ and $g_2$,  the algorithm performs as follows.
\begin{enumerate}
\item Search an injective node mapping $f_{s}$ between $g_1$ and $g_2$ such that $\ns(g_1) \sqsubseteq \ens(g_2)$;

\item Test whether $f_{s}(\es(g_1)) \sqsubseteq \es(g_2)$.
If yes, the algorithm terminates and returns $g_1 \subseteq_t g_2$; 
otherwise, the algorithm searches next qualified node mapping.
If such a node mapping exists, repeat step 1 and 2;
otherwise, the algorithm terminates with $g_1 \nsubseteq_t g_2$.
\end{enumerate}

Although we can perform subsequence tests in linear time, many possible node mappings may exist. Among all the possible mappings, a large amount of them are false mappings, which are not injective. To further improve the speed, we adapt existing pruning techniques for subsequence matching to temporal subgraph tests. The idea is to identify false mappings as early as possible by leveraging node labels, local neighborhood information, and processed prefixes as conditions to prune unpromising search branches. We detail these pruning techniques in Appendix J. 

\subsection{Residual Graph Set Equivalence}
\label{sec:pruning:status}

In this section, we discuss how to efficiently test equivalent residual graph sets by leveraging temporal information in graphs.
A naive approach applies a linear scan algorithm.
Since residual graph set equivalence tests are frequently employed by subgraph and supergraph pruning, repeated linear scans causes significant overhead and suffers poor efficiency.
%Next, we show how temporal information can help us test equivalent residual graph sets in constant time.

Let $g_1$ and $g_2$ be temporal graph patterns.
Consider $G'_1$ and $G'_2$ are the matches of $g_1$ and $g_2$ in $G$, respectively.
Since edges in temporal graphs are totally ordered, we can derive the following result:
$R(G, G'_1)$ is equivalent to $R(G, G'_2)$ if and only if $|R(G, G'_1)| = |R(G, G'_2)|$.
Thus, we can efficiently conduct residual graph set equivalence tests as follows.

\begin{lemma}
Given temporal graph patterns $g_1$ and $g_2$ with $g_1 \subseteq_t g_2$, and a set of graphs $\mathbb{G}$,
$\mathbb{R}(\mathbb{G}, g_1) = \mathbb{R}(\mathbb{G}, g_2)$ if and only if $I(\mathbb{G}, g_1) = I(\mathbb{G}, g_2)$,
where 
\[
I(\mathbb{G}, g_i) = \sum_{R(G, G') \in \mathbb{R}(\mathbb{G}, g_i)} |R(G, G')|.
\]
\label{lm:residual-equivalence}
\end{lemma}

\begin{proof}
The proof is detailed in Appendix K.
\end{proof}
\textbf{Remark}. 
We only need to pre-compute $I(\mathbb{G}, g)$ once by a linear scan over $\mathbb{R}(\mathbb{G}, g)$,
and the following residual graph set equivalence tests are performed in constant time.

\section{Discussion: Concurrent Edges}
\label{sec:concurrent}

In a system with parallelism and concurrency, its monitoring data may generate concurrent edges (\ie edges sharing identical timestamps). In the following, we discuss how our technique can handle such cases.

First, we can extend \tgmine to mine patterns of concurrent edges by the following modifications.
\begin{itemize}
\item For temporal graph representation, instead of using a sequence of edges, we use a sequence of concurrent subgraphs, each of which includes all the edges sharing identical timestamps.

\item In terms of pattern growth, if the added edge has a larger timestamp compared with the last added edge, follow the growth algorithm in \tgmine; if the added edge share the same timestamp with the last added edge, follow the growth algorithm in \gspan~\cite{yan2002gspan}; and we never add edges with smaller timestamps. To avoid repeated patterns, each concurrent subgraph is encoded by canonical labeling.

\item For subgraph tests in pruning, we need to replace node matching based on labels with concurrent subgraph matching based on subgraph isomorphism, and ensure node mappings are injective. 
\end{itemize}
Note that the computation complexity in the extended \tgmine will be increased, as the costly subgraph isomorphism is unavoidable for dealing with the non-temporal graphs formed by concurrent edges.

Second, instead of modifying the mining algorithm, we can transform concurrent edges into total-ordered edges. Despite of the existence of concurrency, data collectors can sequentialize concurrent events based on pre-defined policies~\cite{wang2010algorithmic, zong2014kdd} (\eg randomly assigning a total time order for concurrent edges). In this way, we use the monitoring data with an artificial total order to approximate the original data, and apply \tgmine without modification. When there are a small portion of concurrent edges with minor accuracy loss, this method benefits the efficiency of \tgmine.
\section{Experimental Results}
\label{sec:experiment}

In this section, we evaluate the proposed algorithms for behavior query discovery using real system activity data (\ie syscall logs). In particular, we focus on two aspects: (1) the effectiveness of the behavior queries found by our algorithms, and (2) the efficiency of the proposed algorithms.  

\subsection{Setup}
\label{sec:experiment:setup}

We start with the description for the datasets investigated in our experimental study.

\begin{table}[tbh]
\center
\scalebox{0.72}{
\begin{tabular}{| c | c | c | c | c |}
\hline
Behavior & Avg. $\#$nodes & Avg. $\#$edges & Total $\#$labels &  Size  \\
\hline \hline
bzip2-decompress & 11 & 12 & 15 & small \\
gzip-decompress & 10 & 12 & 7 & small \\
wget-download & 33 & 40 & 92 & small \\
ftp-download & 30 & 61 & 39 & small \\
scp-download & 50 & 106 & 68 & medium \\
gcc-compile & 65 & 122 & 94 & medium \\
g++-compile & 67 & 117 & 100 & medium \\
ftpd-login & 28 & 103 & 119 & medium \\
ssh-login & 66 & 161 & 94 & medium \\
sshd-login & 281 & 730 & 269 & large \\
apt-get-update & 209 & 994 & 203 & large \\
apt-get-install & 1006 & 1879 & 272 & large \\
background & 172 & 749 & 9065 & -- \\
\hline
\end{tabular}
}
\caption{Statistics in training data}
\label{table:data}
\end{table}

\textbf{Training data}.
We collect 13 datasets with in total $11,200$ temporal graphs to mine behavior queries for 12 different behaviors. In total, it contains $1,905,621$ nodes and $7,923,788$ edges derived from the syscall logs including behaviors of interest and background system activities.

We focus on 12 behaviors as representatives for the basic behaviors that have drawn attention in cybersecurity study~\cite{sshattack, bayer2009view, invernizzi2014nazca, yin2007panorama}. For each behavior, we collect 100 temporal graphs based on the syscall logs generated from 100 independent executions of the behavior, as we find the sample size of 100 already achieves good precision (97$\%$) and recall (91$\%$). For background system activities, $10,000$ temporal graphs are sampled from $7$ days' syscall logs generated by a server without performing any target behaviors.

The statistics of the training data are shown in Table~\ref{table:data}. Although the size of a single temporal graph is not large, we find discriminative patterns within the data can involve up to 45 edges, which is a large number for pattern mining problems because of the exponential number of sub-patterns. 

To evaluate how the effectiveness and efficiency are affected by different amounts of training data, we vary the amount of used training data from $0.01$ to $1.0$, where $0.01$ means 1$\%$ of training data are used for behavior query discovery, and $1.0$ means we consider all the data. 

In addition, we generate synthetic datasets to evaluate the scalability of {\tgmine}. The synthetic datasets are created based on the training data: we generate datasets {\syn}-2, {\syn}-4, {\syn}-6, {\syn}-8, and {\syn}-10 by replicating each graph in the training data 2, 4, 6, 8, and 10 times, respectively.  

\textbf{Test data}. 
Test data are used to evaluate the accuracy of the behavior queries found by our algorithms. They are obtained from an independent data collection process: we collect another seven days' syscall log data, which forms a large temporal graph with $2,352,204$ nodes and $15,035,423$ edges. The test data contain $10,000$ behavior instances of the 12 target behaviors. Note that our focus in this paper is query formulation instead of pattern query processing. Based on the patterns discovered from training data, we formulate behavior queries, and search their existence from the test data by existing techniques~\cite{zong2014cloud}.

% nodes: 2,352,204; edges: 15,035,423

More details about training/test data collection are presented in Appendix L. 

\textbf{Implementation}.
For effectiveness, we consider \static and \nodeset as baselines. 
(1) \static employs non-temporal graph patterns for behavior query discovery. In particular, we remove all the temporal information in the training data, apply existing algorithms~\cite{jin2010gaia} to mine discriminative non-temporal graph patterns for each behavior, and use the discovered patterns to formulate non-temporal behavior queries. 
(2) \nodeset searches behavior instances by keyword queries using a set of discriminative node labels. The discriminativeness of a node label is measured by the same score function $F(x, y)$ for temporal graph patterns. Top-$k$ discriminative node labels are selected for a query of $k$ node labels. A match of a query is a set of $k$ nodes, where its node label set is identical to the node label set specified in the query, and its spanned time interval is no longer than the longest observed lifetime of the target behavior.  

For efficiency, we implement five baseline algorithms to demonstrate the contributions of each component in {\tgmine}. All the baselines apply the proposed pattern growth algorithm and the naive pruning condition stated in Section~\ref{sec:pruning:naive}.
(1) \subprune employs the pruning condition in Lemma~\ref{lm:subgraph}.
(2) \supprune considers the pruning condition in Proposition~\ref{prop:supergraph}.
(3) \prunegi uses all the pruning conditions but applies a graph index based algorithm for temporal subgraph tests.
In particular, we index one-edge substructures, and use efficient algorithms to join partial matches into full matches~\cite{zong2014cloud}.
(4) \vfprune considers all the pruning conditions but performs temporal subgraph tests by a modified VF2 algorithm~\cite{cordella2004sub}.
(5) \lscan uses all the pruning conditions but performs residual graph set equivalence tests via a linear scan algorithm.

In addition, we employ information gain, G-test~\cite{yan2008mining}, as well as the function $F(x, y) = \log{(x / (y + \epsilon))}$ ($\epsilon$ is set to  $10^{-6}$) adopted in~\cite{jin2010gaia} as discriminative score functions. In our experiment, we find these score functions deliver a common set of discriminative patterns.

All the algorithms are implemented in C++ with GCC 4.8.2, and all the experiments are performed on a server with Ubuntu 14.04, powered by an Intel Core i7-2620M 2.7GHz CPU and 32GB of RAM. Each experiment is repeated $10$ times, and their average results are presented.

\subsection{Effectiveness of Behavior Queries}
\label{sec:experiment:effectiveness}

We evaluate the effectiveness of \tgmine from three aspects.
(1) For different behaviors, how accurately can behavior queries suggested by \tgmine search behavior instances from system activity data?
(2) How does query accuracy vary when pattern size in queries changes?
(3) How does the amount of training data affect query accuracy?

Precision and recall are used as the metrics to evaluate the accuracy. Given a target behavior and its behavior query, a match of this behavior query is called an identified instance. An identified instance is correct, if the time interval during which the match happened is fully contained in a time interval during which one of the true behavior instances was under execution. A behavior instance is discovered, if the behavior query can return at least one correct identified instance with respect to this behavior instance. The precision and recall of a behavior query are defined as follows.
\begin{align*}
& \text{precision} = \frac{\#\text{correctly identified instances}}{\text{total }\#\text{identified instances}}, \\
& \text{recall} = \frac{\#\text{discovered instances}}{\#\text{behavior instances}}.
\end{align*}

Note that when \tgmine returns multiple discriminative patterns that have the same highest discriminative score,
the returned patterns are further ranked by a score function based on domain knowledge.
From all the discriminative patterns,
top-$5$ patterns are used to build behavior queries. 
The details about the domain knowledge based score function is discussed in Appendix M.

\begin{figure}[tbh]
\center
\includegraphics[scale=0.45]{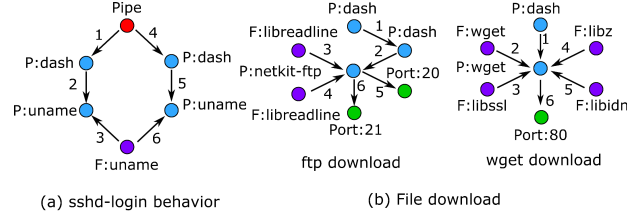}
\caption{Discovered discriminative patterns}
\label{fig:exp:effectiveness:case}
\end{figure}

Figure~\ref{fig:exp:effectiveness:case} illustrates a few discovered discriminative patterns.
For sshd-login behavior, the unique interaction pattern among system entities makes it accurate for search purpose.
Note that this pattern does not include any node with ``sshd'' in the label.
This indicates that keyword-based techniques that simply use application names as keywords (\eg sshd) cannot find such highly accurate patterns.
For file download behaviors, it is the distinct patterns of how to access libraries and sockets that differentiate wget-based download from ftp-based download.

\begin{table}[tbh]
\center
\scalebox{0.70}{
\begin{tabular}{| c | c | c | c | c | c | c |}
\hline
Metric  & \multicolumn{3}{c|}{Precision ($\%$)} & \multicolumn{3}{c|}{Recall ($\%$)}  \\
\hline
Algorithm & \nodeset & \static & \tgmine & \nodeset & \static & \tgmine \\
\hline \hline
bzip2-decompress & 100 & 100 & 100 & 100 & 100 & 100 \\
gzip-decompress & 96.6 & 100 & 100 & 100 & 100 & 100 \\
wget-download & 96.5 & 100 & 100 & 93.6 & 93.4 & 93.4 \\
ftp-download & 100 & 100 & 100 & 100 & 96.1 & 96.1 \\
scp-download & 13.8 & 59.4 & \textbf{100} & 11.2 & 91.3 & 91.3 \\
gcc-compile & 69.7 & 81.2 & \textbf{94.3} & 89.2 & 89.4 & 87.6 \\
g++-compile & 73.4 & 91.3 & \textbf{95.2} & 84.5 & 85.3 & 85.3 \\
ftpd-login & 76.6 & 81.8 & \textbf{94.1} & 100 & 89.7 & 86.8 \\
ssh-login & 33.8 & 64.3 & \textbf{93.9} & 78.7 & 87.2 & 85.9 \\
sshd-login & 43.4 & 59.6 & \textbf{99.9} & 99.8 & 99.9 & 99.9 \\
apt-get-update & 50.3 & 79.3 & \textbf{95.9} & 47.6 & 84.5 & 82.4 \\
apt-get-install & 68.3 & 81.7 & \textbf{95.7} & 35.6 & 86.3 & 83.9 \\
\hline \hline
Average & 68.5 & 83.2 & \textbf{97.4} & 78.4 & 91.9 & 91.1 \\
\hline
\end{tabular}
}
\caption{Query accuracy on different behaviors}
\label{table:query-accuracy}
\end{table}

Table~\ref{table:query-accuracy} shows the precision and recall of behavior queries on all the 12 behaviors.
The size (\ie the number of edges) of the behavior queries suggested by \tgmine and \static is fixed as $6$, using all the training data.
\nodeset employs the top-$6$ discriminative node labels to query each behavior.
First, the behavior queries suggested by \tgmine accurately discover behavior instances.
Over all the behaviors, its average precision and recall are $97.4\%$ and $91.1\%$, respectively.
Second, the queries provided by \static only achieve $83.2\%$ of precision, suffering significantly higher false positive rate.
This result indicates the importance of temporal information in searching system behaviors. 
Third, the queries suggested by \tgmine consistently outperform the queries formed by \nodeset.
These results confirm that temporal graph patterns discovered by \tgmine provide high-quality skeletons to formulate accurate behavior queries.

\begin{figure}[tbh]
\center
\includegraphics[scale=0.7]{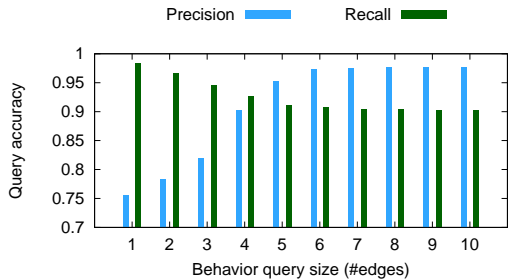}
\caption{Query accuracy with different query sizes}
\label{fig:exp:effectiveness:qsize}
\end{figure}

Figure~\ref{fig:exp:effectiveness:qsize} demonstrates how the precision and recall of behavior queries suggested by \tgmine vary while their query size ranges from $1$ to $10$.
All the training data were used in this experiment,
and the average precision and recall over all the behaviors are reported.
First, when the query size increases, the precision increases, but the recall decreases.
In general, increasing behavior query size improves query precision at the cost of a slightly higher false negative rate.
Second, when the query size goes beyond $6$, we observe little improvement on precision or loss on recall.

\begin{figure}[tbh]
\center
\includegraphics[scale=0.7]{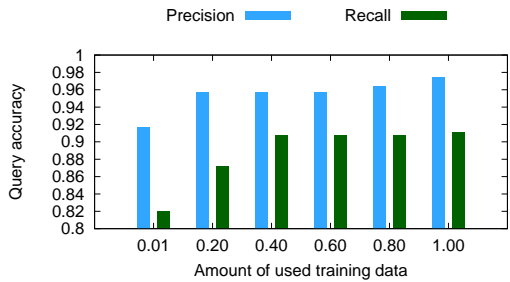}
\caption{Query accuracy with different amounts of used training data}
\label{fig:exp:effectiveness:tdsize}
\end{figure}

Figure~\ref{fig:exp:effectiveness:tdsize} suggests how precision and recall of behavior queries vary as the amount of used training data is changed from $0.01$ to $1.0$.
Note that behavior query size is fixed as $6$,
and the average precision and recall over all the behaviors are presented.
First, more training data bring higher precision and recall.
For example, when the amount of used training data increases from $0.01$ to $1.0$,
the precision of the constructed behavior queries increase from $91\%$ to $97\%$.
Second, when the amount of used training data increases, we observe a diminishing return for both precision and recall.

%===========================================================================================
%===========================================================================================
%===========================================================================================

\subsection{Efficiency in Behavior Query Discovery}
\label{sec:experiment:efficiency}

The efficiency of \tgmine is evaluated from two aspects. (1) How is the efficiency of \tgmine compared with baseline algorithms? (2) How is the efficiency of \tgmine affected by different amounts of used training data? Note that behavior query discovery is an offline step: we only need to mine patterns once, and then use the patterns to formulate queries serving online search demands.

Response time is used as the metric to evaluate the efficiency. Given input temporal graph sets, the response time of an algorithm is the amount of time the algorithm spends in mining all discriminative patterns.

\begin{figure*}[tbh!]
\center
\subfigure[Small behavior traces]{
\includegraphics[scale=0.73]{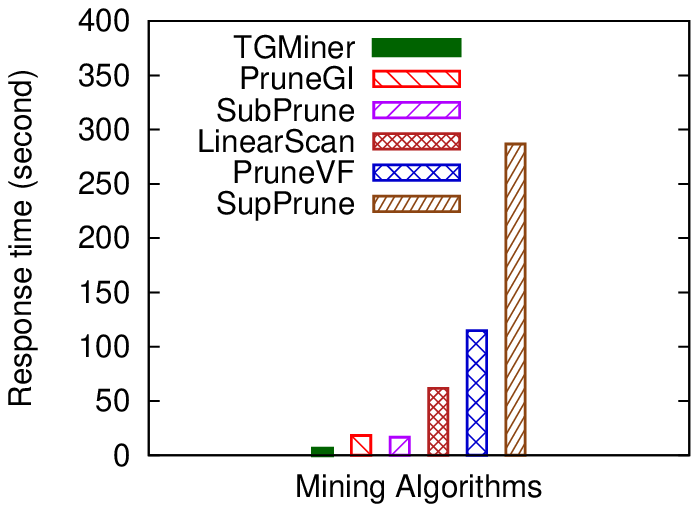}
\label{fig:exp:efficiency:small}
}
%\hfill
\subfigure[Medium behavior traces]{
\includegraphics[scale=0.73]{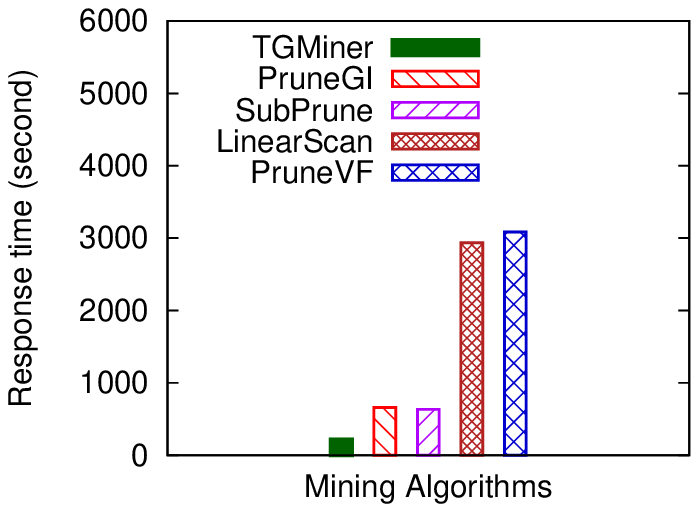}
\label{fig:exp:efficiency:medium}
}
%\hfill
\subfigure[Large behavior traces]{
\includegraphics[scale=0.73]{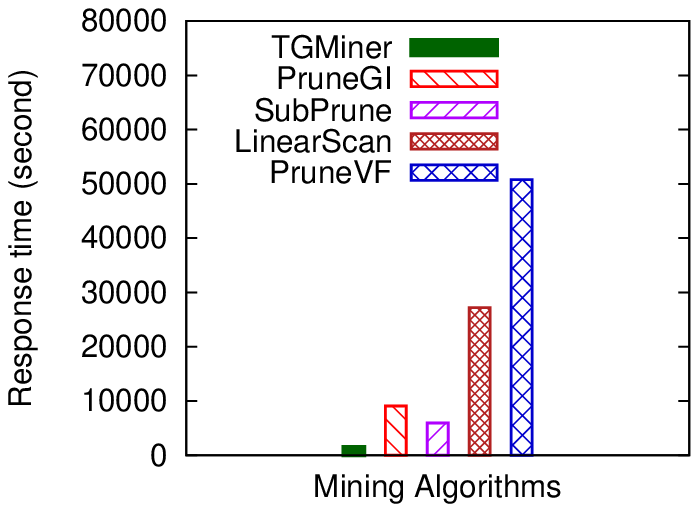}
\label{fig:exp:efficiency:large}
}
%\hfill
\caption{Response time for behaviors of different sizes}
\label{fig:exp:efficiency:qsize}
\end{figure*}

Figure~\ref{fig:exp:efficiency:qsize} demonstrates response time of all algorithms over small, medium, and large size behaviors as categorized in Table~\ref{table:data}. All the training data are used in this experiment, and the discovered patterns have up to 45 edges.

First, \tgmine consistently outperforms the baseline algorithms over all the target behaviors. \tgmine is up to $50$ and $4$ times faster than \subprune and \supprune. \supprune cannot finish the mining tasks for medium and large behaviors within 2 days. We notice most of the pruning opportunities come from subgraph pruning, while supergraph pruning brings additional performance improvement.

Second, \tgmine performs up to $6$, $17$, and $32$ times faster than {\prunegi}, {\lscan},  and {\vfprune}, respectively. \prunegi has to frequently build graph indexes for each discovered patterns during the whole mining process, which involves high overhead. Indeed, graph indexing is more suitable for querying large graphs, where we can build graph indexes offline. In our case, a light-weighted temporal subgraph test algorithm performs better. In sum, the performance improvement highlights the importance of minimizing the overhead in temporal subgraph tests and residual graph set equivalence tests.

Third, for small, medium, and large behaviors, \tgmine can complete mining tasks within 6 seconds, 4 minutes, and  26 minutes, respectively. We also noticed that \tgmine mines all discriminative patterns with no more than 6 edges within one minute for all the behaviors, while 6-edge patterns can achieve good search accuracy as shown before.

\begin{figure}[tbh]
\center
\includegraphics[scale=0.7]{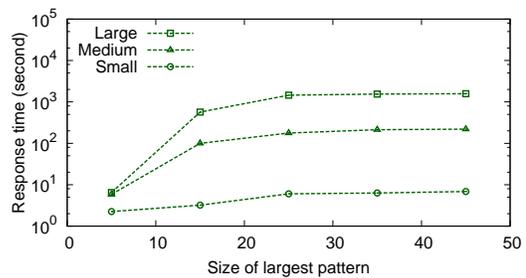}
\caption{Response time by varying the size of the largest patterns that are allowed to explore}
\label{fig:exp:efficiency:qsize-all}
\end{figure}

Figure~\ref{fig:exp:efficiency:qsize-all} presents the response time of \tgmine as the size of the largest patterns that are allowed to explore is set to $5$, $15$, $25$, $35$, and $45$. When the size increases, the response time of \tgmine increases. When the size is set as 5, \tgmine can finish the mining tasks within $10$ seconds for all the behaviors,.

\begin{table}[tbh]
\center
\scalebox{1.0}{
\begin{tabular}{| c | c | c | c |}
\hline
Pruning Condition & Small & Medium & Large  \\
\hline \hline
Subgraph pruning & $71.8\%$ & $61.0\%$ & $62.2\%$ \\
Supergraph pruning & $1.1\%$ & $8.3\%$ & $4.2\%$ \\
\hline
\end{tabular}
}
\caption{Empirical probabilities that pruning conditions are triggered on behaviors of different sizes}
\label{table:em-prob}
\end{table}

Table~\ref{table:em-prob} shows the empirical probabilities that subgraph and supergraph pruning are triggered when \tgmine is processing a pattern for behaviors of different sizes. The high trigger rate of subgraph pruning is consistent over all the behaviors, which explains its high pruning power.

\begin{figure}[tbh]
\center
\includegraphics[scale=0.7]{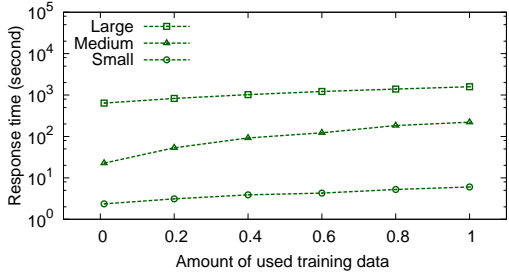}
\caption{Response time by varying the amount of used training data}
\label{fig:exp:efficiency:tdsize}
\end{figure}

Figure~\ref{fig:exp:efficiency:tdsize} shows how the response time of \tgmine is affected by using different amounts of training data. Ranging from $0.01$ to $1.0$, the response time of \tgmine increases linearly when more training data is considered. The scalability test on synthetic datasets (detailed in Appendix N) also shows that the response time of \tgmine linearly scales with the size of training data. From the training data with up to 20M nodes and 80M edges, \tgmine can mine all discriminative patterns of up to 45 edges within 3 hours.

\subsection{Summary}
\label{sec:experiment:summary}

The experimental results are summarized as follows.
First, behavior queries discovered by \tgmine achieve high precision ($97\%$) and recall ($91\%$), while the baseline algorithms \static and \nodeset suffers poor accuracy.
Second, \tgmine consistently outperforms the baseline algorithms in terms of efficiency, and is up to $6$, $17$, and $32$  times faster than {\prunegi}, {\lscan},  and {\vfprune}, respectively.

\section{Related Work}
\label{sec:related}

\subsection{Discriminative Graph Pattern Mining}
\label{sec:related:dpm}

Discriminative graph pattern mining is one of the feature selection methods that is widely applied in a variety of graph classification tasks~\cite{ranu2009graphsig, thoma2009near, yan2008mining}.

In general, two directions have been investigated to speed up mining discriminative non-temporal graph patterns. One direction is to find discriminative patterns early such that unpromising search branches can be pruned early~\cite{yan2008mining}. The other direction relies on approximate search that finds good-enough graph patterns~\cite{jin2010gaia}. In addition, a few studies~\cite{qin2011evolution, ranu2013mining, wackersreuther2010frequent} focus on mining graph patterns from a sequence of graph snapshots.

It is difficult to extend existing work on non-temporal graphs to mine temporal graph patterns. The key problem is how to deal with timestamps in the mining process. 

One possibility is to ignore timestamps: mine discriminative non-temporal patterns by existing approaches, and then use timestamps to find discriminative temporal patterns from the non-temporal patterns. This method has two drawbacks. First, since canonical labeling on non-temporal graphs~\cite{jin2010gaia, yan2002gspan} have difficulties in dealing with multi-edges, we have to collapse multi-edges into a single edge. In this way, the final result will be partial, as it excludes patterns with multi-edges. Second, a large number of temporal patterns may share the same non-temporal patterns, and a discriminative non-temporal pattern may result in no discriminative temporal pattern. The redundancy in non-temporal patterns will bring potential scalability problems.

Another possibility is to consider timestamps as labels. The temporal patterns discussed in this paper focus on the temporal order instead of exact timestamps. Therefore, it is difficult to mine the desired patterns by the approaches for non-temporal patterns.

Our work is different from existing works. First, unlike existing studies that mainly deal with non-temporal graphs, we propose a solution to temporal graphs. Second, compared with existing works on graph snapshots, we study more flexible temporal graph patterns. The graph patterns defined in~\cite{qin2011evolution, ranu2013mining, wackersreuther2010frequent} are too rigid to support applications in system management.

\subsection{Temporal Graph Management}
\label{sec:related:tgman}

Recent studies on temporal graphs strive to develop cost-effective solutions to management problems.

There have been a few works on developing more efficient index-free algorithms for time-dependent shortest-path~\cite{yang2014finding}, minimum temporal path~\cite{wu2014path}, and anomaly detection~\cite{Sricharan2014}. 

Graph indexing, compression, and partitioning have been proposed to accelerate query processing for subgraph matching~\cite{zong2014cloud}, reachability~\cite{Zhu2014}, community searching~\cite{huang2014querying}, and neighborhood aggregation~\cite{Mondal2014}.
Incremental computation has been investigated for queries including shortest-path~\cite{ren2011querying}, SimRank~\cite{yu2014fast}, and cluster searching~\cite{lee2014incremental}.
In addition, Gao et al.~\cite{gao2014continuous} exploited distributed computation to monitor subgraph matching queries.

Data storage is a critical component in temporal graph management. 
Grace~\cite{prabhakaran2012managing} examined graph update transactions that incorporate dynamic changes into the latest graph snapshot.
Chronos~\cite{hant2014chronos} investigated temporal data locality for query processing.
Efficient ways to store and retrieve graph snapshots have been studied in~\cite{khurana2013efficient, labouseur2013scalable}.

%Kineograph~\cite{cheng2012kineograph} addressed the consistency problem on storing graph snapshots,

Unlike these works, our focus is to develop cost-effective algorithms that help users formulate meaningful temporal graph queries, bridging the gap between users' knowledge and temporal graph data.

%===========================================================================================
%===========================================================================================
%===========================================================================================

\newpage
\section{Conclusion}
\label{sec:conclusion}

Computer system monitoring generates huge temporal graphs that record the interaction of system entities.
While these graphs are promising for system experts to query behaviors for system management, 
it is also difficult for them to compose queries as it involves many tedious low-level system entities.
We formulated this query formulation problem as a discriminative temporal graph mining problem,
and introduced \tgmine to mine discriminative patterns that can be taken as query templates for building more complex queries. 
\tgmine leverages temporal information in graphs to enable efficient pattern space exploration and prune unpromising search branches. 
Experimental results on real system data show that \tgmine is 6-32 times faster than baseline methods.  
Moreover, the discovered patterns were verified by system experts: they achieved high precision ($97\%$) and recall ($91\%$).

\textbf{Acknowledgement}. We would like to thank the anonymous reviewers for the helpful comments on earlier versions of the paper. 
This research was partially supported by the Army Research Laboratory under cooperative agreements W911NF-09-2-0053 (NS-CTA), NSF IIS-1219254, and NSF IIS-0954125.
The views and conclusions contained herein are those of the authors and should not be interpreted as representing the official policies, either expressed or implied, of the Army Research Laboratory or the U.S. Government. The U.S. Government is authorized to reproduce and distribute reprints for Government purposes notwithstanding any copyright notice herein.

\bibliographystyle{abbrv}
\begin{small}
\bibliography{ref}
\end{small}

\appendix

\section{Proof of Lemma 1}
\begin{proof}
Without loss of generality, we assume $g_1 = (V_1, E_1, A_1, T_1)$ and $g_2 = (V_2, E_2, A_2, T_2)$.

First, we prove $\tau: T_1 \rightarrow T_2$ is unique.
Since $g_1$ and $g_2$ are temporal graph patterns, 
we have $\forall (u_1, v_1, t_1) \in E_1$, $1 \leq t_1 \leq |E_1|$ and $\forall (u_2, v_2, t_2) \in E_2$, $1 \leq t_2 \leq |E_2|$.
Because $g_1 =_t g_2$ and $|E_1| = |E_2|$,
we can derive $(u_1, v_1, t_1) \in E_1$ matches $(u_2, v_2, t_2) \in E_2$ only if $t_1 = t_2$ (in order to preserve total edge order).
Thus, the uniqueness of $\tau$ is proved.

Second, we prove $f: V_1 \rightarrow V_2$ is unique.
Since $\tau$ is unique, the edge mapping between $g_1$ and $g_2$ is unique,
and then we can derive the node mapping $f$ is also unique. 

Therefore, we have proved the uniqueness of $f$ and $\tau$.
\end{proof}

\section{Proof of Lemma 2}
\begin{proof}
We first present the linear algorithm, and then prove its correctness. Suppose $g_1 = (V_1, E_1, A_1, T_1)$, $g_2 = (V_2, E_2, A_2, T_2)$, $|V_1| = |V_2|$, and $|E_1| = |E_2|$.

\textbf{Algorithm}. We conduct a linear scan over edges in $g_1$. 
For each edge $(u_1, v_1, t) \in E_1$, we locate the edge $(u_2, v_2, t) \in E_2$.
If such an edge exists, we verify if the mapping from $u_1$ to $u_2$  and the mapping from $v_1$ to $v_2$ are still one-to-one.
If both are, we confirm $(u_1, v_1, t)$ matches $(u_2, v_2, t)$.
Then $g_1 =_t g_2$ only if all the edges in $g_1$ find their matches in $g_2$.

\textbf{Correctness}. We prove the correctness by showing the above algorithm finds two bijective functions $f: V_1 \rightarrow V_2$ and $\tau: T_1 \rightarrow T_2$.  First, since the linear scan follows the unique way to match edge timestamps between $g_1$ and $g_2$ and $|E_1| = |E_2|$, $\tau$ is found and bijective. Second, the verification phase in the above algorithm guarantees the node mapping $f$ is one-to-one; moreover, $f$ is a full mapping because $|E_1| = |E_2|$ and all the nodes in $g_1$ and $g_2$ find their mapping in the verification phase. Therefore, we have proved the correctness of the linear algorithm.
\end{proof}

\section{Proof of Lemma 3}
\begin{proof}
Let $g_1$ and $g_2$ be connected temporal graph patterns with $g_1 \subseteq g_2$.
Suppose the edge sets of $g_1$ and $g_2$ are $E_1$ and $E_2$, respectively.
Then we need to conduct $m = |E_2| - |E_1|$ steps of consecutive growth to grow $g_1$ into another pattern $g'_2$. 
If there exists $g'_2 =_t g_2$, then we can tell it is possible to grow $g_1$ into $g_2$;
otherwise, there is no way to grow $g_1$ to $g_2$.
Next we prove if we can grow $g_1$ into $g_2$, then the $m$ steps of consecutive growth is unique.   

Assume that (1) $s' = \langle e'_{1}, e'_{2}, \cdots, e'_{m} \rangle$ is a sequence of consecutive growth that grows $g_1$ into $g'_2$ with $g'_2 =_t g_2$,   (2) $s'' = \langle e''_{1}, e''_{2}, \cdots, e''_{m} \rangle$ is another sequence of consecutive growth that grows $g_1$ into $g''_2$ with $g''_2 =_t g_2$, and (3) $s'$ is distinct from $s''$ as $\exists (u', v', t) \in s'$ cannot match $(u'', v'', t) \in s''$.
On the one hand, since $g'_2 =_t g_2$ and $g''_2 =_t g_2$, we can infer $g'_2 =_t g''_2$ by the bijective mapping functions.
On the other hand, by the definition of consecutive growth,  the linear scan algorithm from Lemma~\ref{lm:linear-match} will decide $g'_2$ cannot match $g''_2$, since there exists at least one edge from $s'$ that cannot match the edge in $s''$ sharing the same timestamp, 
which contradicts with $g'_2 =_t g''_2$. 
Thus, $s'$ is identical to $s''$, and the $m$ steps of consecutive growth is unique.

In sum, we have proved the correctness of Lemma~\ref{lm:forward}.
\end{proof}

\section{Proof of Theorem 1}
\begin{proof}
We first justify the zero-redundancy in pattern search, and then prove the completeness.

Let $g$ be a connected temporal graph pattern.
By Lemma~\ref{lm:forward}, 
we can tell consecutive growth guarantees there exists a unique way to grow an empty pattern into $g$.
Thus, there is no way to search $g$ more than once.

For the completeness, we establish the proof by mathematical induction.
Let $m$ be the number of edges in a temporal graph pattern.

\textbf{Basis}: we prove the completeness holds for $m = 1$.
Since $1$-edge patterns are formed by distinct edges,
it is easy to see all $1$-edge patterns will be covered.

\textbf{Inductive step}: we prove if the completeness holds for $m = k$, 
then it holds for $m = k + 1$.

Assume the completeness holds for $m = k$.
Then we have the complete set of $k$-edge connected temporal graph patterns $H^{(k)}$.
Assume $g^{(k+1)} = g^{(k)} \cup \{ e\}$ is a connected pattern of $k+1$ edges that is grown from a pattern $g^{(k)}$ of $k$ edges.
Since the three growth options are all possible ways to keep patterns connected during growth,
if $g^{(k+1)}$ cannot be covered by growing patterns in $H^{(k)}$,
it implies $g^{(k)} \notin H^{(k)}$, that is, $g^{(k)}$ is not connected,
which contradicts with the assumption that $g^{(k+1)}$ is connected (T-connected). 
Therefore, the completeness also holds for $m = k+1$.

Since both the basis and inductive step have been performed, by mathematical induction,
the completeness holds.

In sum, we have proved Theorem~\ref{thm:growth}.
\end{proof}

\section{Proof of Proposition 1}
\label{app:propositionx}
\begin{proof}
Let $g_1$ and $g_2$ be temporal graph patterns and $\mathbb{G}$ be a set of temporal graphs.
Assume we have $g_1 \subseteq_t g_2$ with two distinct node mappings $f$ and $f'$, 
and $\mathbb{R}(\mathbb{G}, g_1) = \mathbb{R}(\mathbb{G}, g_2)$. 
For any $G \in \mathbb{G}$ such that $G' \subseteq_t G$ and $G'$ matches $g_2$,
we can find two distincted subgraphs in $G'$ that match $g_1$ because of $f$ and $f'$.
Therefore, $|\mathbb{R}(\mathbb{G}, g_1)| = 2 | \mathbb{R}(\mathbb{G}, g_2) |$,
which contradicts with the fact that $\mathbb{R}(\mathbb{G}, g_1) = \mathbb{R}(\mathbb{G}, g_2)$.
Therefore, we have proved the proposition holds.
\end{proof}

\section{Proof of Lemma 4}
\begin{proof}
Let $g_1$ and $g_2$ be temporal graph patterns, where $g_1$ is discovered before $g_2$ and they satisfy the conditions in subgraph pruning.

First, since the conditions in subgraph pruning are satisfied, 
we can derive the following facts: (1) $\freq(\mathbb{G}_p, g_2) = \freq(\mathbb{G}_p, g_1)$ and (2) pattern growth in $g_1$'s branch will \emph{never} touch the nodes that cannot map to any nodes in $g_2$ as $\mathbb{L}(\mathbb{G}_p, g_2) \cap L_{g_1 \setminus g_2} = \emptyset$.

Second, we prove none of the patterns in $g_2$'s branch will be the most discriminative, if the conditions in subgraph pruning are satisfied and none of the patterns in $g_1$'s branch is the most discriminative.
Assume there exists a pattern $g'_2$ whose discriminative score is no less than $F^*$ and $s$ is the sequence of consecutive growth that grows $g_2$ into $g'_2$.
Since no pattern growth in $g_1$'s branch will touch the nodes that cannot map to any nodes in $g_2$, we can safely claim $s$ also indicates a valid sequence of consecutive growth (with some timestamp shift) that grows $g_1$ into $g'_1$ by Proposition~\ref{prop:proposition}.
By $\freq(\mathbb{G}_p, g_2) = \freq(\mathbb{G}_p, g_1)$ and $\mathbb{R}(\mathbb{G}_p, g_2) = \mathbb{R}(\mathbb{G}_p, g_1)$,
we can infer $\freq(\mathbb{G}_p, g'_2) = \freq(\mathbb{G}_p, g'_1)$. 
It is not hard to see $g'_2 \subseteq_t g'_1$. Thus, we have $\freq(\mathbb{G}_n, g'_2) \geq \freq(\mathbb{G}_n, g'_1)$.
Therefore, we can infer $F(g'_2) \leq F(g'_1)$, which means $g'_1$ is one of the most discriminative patterns which contradicts with the condition that none of the patterns in $g_1$'s branch is the most discriminative.
Therefore, we can claim any patterns in $g_2$'s branch will have discriminative score less than $F^*$, and the branch can be safely pruned.

In all, we have proved Lemma~\ref{lm:subgraph} holds.
\end{proof}

\section{Proof of Proposition 2}
\label{app:proposition1}

\begin{proof}
Let $g_1$ and $g_2$ be temporal graph patterns, where $g_1$ is discovered before $g_2$ and they satisfy the conditions in supergraph pruning.

First, given the conditions $g_2 \supseteq_t g_1$, 
$\mathbb{R}(\mathbb{G}_p, g_2) = \mathbb{R}(\mathbb{G}_p, g_1)$,
and $\mathbb{R}(\mathbb{G}_n, g_2) = \mathbb{R}(\mathbb{G}_n, g_1)$,
we can derive the following properties,
\begin{enumerate}

\item $\freq(\mathbb{G}_p, g_1) = \freq(\mathbb{G}_p, g_2)$;

\item $\freq(\mathbb{G}_n, g_1) = \freq(\mathbb{G}_n, g_2)$.

\end{enumerate}

Second, we prove none of patterns in $g_2$'s branch will be the most discriminative, if the conditions in supergraph pruning are satisfied and none of patterns in $g_1$'s branch is the most discriminative.
Assume there exsits a pattern $g'_2$ whose discriminative score is no less than $F^*$ and $s$ is the sequence of consecutive growth that grows $g_2$ into $g'_2$.
Since $g_1$ and $g_2$ have the same number of nodes, we can safely claim $s$ is also a valid sequence of consecutive growth (with some timestamp shift) that grows $g_1$ into $g'_1$ by Proposition~\ref{prop:proposition}.
By $\freq(\mathbb{G}_p, g_2) = \freq(\mathbb{G}_p, g_1)$ and $\mathbb{R}(\mathbb{G}_p, g_2) = \mathbb{R}(\mathbb{G}_p, g_1)$, we can infer $\freq(\mathbb{G}_p, g'_2) = \freq(\mathbb{G}_p, g'_1)$. Similarly, we can infer $\freq(\mathbb{G}_n, g'_2) = \freq(\mathbb{G}_n, g'_1)$. Thus, we have $F(g'_2) = F(g'_1)$, which means $g'_1$ is one of the most discriminative patterns which contradicts with the condition that none of patterns in $g_1$'s branch is the most discriminative. Therefore, we can claim any patterns in $g_2$'s branch will have discriminative score less than $F^*$, and the branch can be safely pruned.

In sum, we have proved the correctness of this Proposition.
\label{proof:proposition1}
\end{proof}

\section{Proof of Proposition 3}
\label{app:proposition2}

\begin{proof}
We prove temporal subgraph test problem is NP-hard and NP.

First, we prove the NP-hardness by reducing clique problem~\cite{michael1979computers} to temporal subgraph test problem. 
Given a non-temporal graph $G = (V, E)$ where $V$ and $E$ are node and edge sets,
the decision problem of clique problem is to decide whether there exists a graph $G' = (V', E')$ such that $G'$ is a subgraph of $G$,
$G'$ is a clique, and $|V'| \geq k$.
We transform a non-temporal graph $G = (V, E)$ into a temporal graph $G_t = (V_t, E_t, T)$,
where $V_t$ is a node set, $E \subseteq V_t \times V_t \times T$ is a temporal edge set,
and $T \subseteq \mathbb{N}$ is a set of non-negative integers indicating possible timestamps on edges. 
The transformation is conducted in the following steps.
\begin{enumerate*}

\item We set $V_t = V$.
$\forall v \in V_t$, a unique node ID is assigned, denoted as $id(v)$, 
such that $\forall u, v \in V_t$, either $id(u) > id(v)$ or $id(u) < id(v)$.

\item We transform the undirected edge set $E$ into a directed edge set $E_d$ 
such that $\forall (u ,v) \in E$, $(u, v) \in E_d$ and $(v, u) \in E_d$.

\item We assign timestamps to edges in $E_d$ and form a temporal edge set $E_t$ as follows:
$\forall (u ,v) \in E_d$,
$(u, v, t) \in E_t$ where the timestamp $t = |V|^{id(u)} + id(v)$.
The way we assign timestamps to edges has two implications:
(1) $\forall (u_1, v_1, t_1), (u_2, v_2, t_2) \in E_t$,
if $id(u_1) < id(u_2)$, then $t_1 < t_2$;
and (2) $\forall (u, v_1, t_1), (u, v_2, t_2) \in E_t$,
if $id(v_1) < id(v_2)$, then $t_1 < t_2$.
\end{enumerate*}

Following the same steps, 
a clique graph $G' = (V', E')$ with $|V'| > k$ is transformed into a temporal graph ${G'}_t = ({V'}_t, {E'}_t, T')$.
It is not hard to see $G$ contains a clique of size no less than $k$ if and only if $G'_t \subseteq G_t$.
Since the reduction is performed in polynomial time, 
temporal subgraph test problem is at least as hard as clique problem.
Therefore, we have proved it is NP-hard. 

Second, temporal subgraph test problem is NP.
Let $G_1$ and $G_2$ be two temporal graphs. 
Given the two injective functions $f$ and $\tau$ between $G_1$ and $G_2$,
we can verify whether $G_1 \subseteq_t G_2$ in polynomial time.
Therefore, we have proved it is NP.

In sum, we have proved temporal subgraph test problem is NP-complete.
\label{proof:proposition2}
\end{proof}

\section{Proof for Lemma 5}
\label{app:subseq-subgraph}

\begin{proof}
We prove the sufficiency and necessity as follows.

First, we prove if $g_1 \subseteq_t g_2$,
there exists $\ns(g_1) \sqsubseteq \ens(g_2)$ with an injective node mapping $f'$ such that $f'(\es(g_1)) \sqsubseteq \es(g_2)$.
Let $m$ be the number of edges in $g_1$.
We establish the proof by mathematical induction.
\begin{itemize}
\item Basis: we prove the statement holds for $m = 1$.
If $g_1 \subseteq_t g_2$,
we can find $\ns(g_1) \sqsubseteq \ens(g_2)$, 
since a source node always appears earlier than a destination node in an enhanced node sequence,
and the resulting node mapping $f'$ satisfies $f'(\es(g_1)) \sqsubseteq \es(g_2)$.

\item Inductive step: we prove if the statement holds for $m = k$, it holds for $m = k+1$.
Assume the statement holds for $m = k$.
There are three growth options to grow a $k$-edge temporal graph pattern to a $(k+1)$-edge temporal graph pattern.
Consider $g_1$ is grown from a $k$-edge pattern $g'_1$.
(1) For forward growth,
$\ns(g_1)$ has one more node (a destination node) compared with $\ns(g'_1)$.
We can find a corresponding node in $\ens(g_2)$ that maps the additional node in $\ns(g_1)$,
since enhanced node sequences always record destination nodes.
Therefore, an injective node mapping is formed.
(2) For backward growth,
$\ns(g_1)$ has one more node $u_1$ (a source node) compared with $\ns(g'_1)$.
Let $(v, w, t)$ be the edge in $g_2$ that matches the last edge in $g'_1$.
The node $u_2$ in $g_2$ that maps $u_1$ cannot be $v$ or $w$;
therefore, $u_2$ will be recorded after $w$ in $\ens(g_2)$ and match $u_1$.
Therefore, an injective node mapping is formed.
(3) For inward growth,
no more node will be added into $\ns(g_1)$ compared with $\ns(g'_1)$.
Therefore, an injective mapping is formed.
For all the cases,
as long as an injective mapping is formed,
we can find $f'(\es(g_1)) \sqsubseteq \es(g_2)$.

Therefore, the statement holds for $m = k+1$.
\end{itemize}
Since both the basis and the inductive step have been performed, by mathematical induction,
the sufficiency holds for all $m > 0$.

Second, we prove if $\ns(g_1) \sqsubseteq \ens(g_2)$ with an injective node mapping $f'$ such that $f'(\es(g_1)) \sqsubseteq \es(g_2)$,
then $g_1 \subseteq_t g_2$. 
Through the node mapping $f'$,
we can construct a temporal graph $g'_2$ in $g_2$ such that $g_1 \subseteq_t g'_2$.
Since $g'_2 \subseteq_t g_2$,
we can conclude $g_1 \subseteq_t g_2$.
  
In sum, Lemma~\ref{lm:subseq} is proved.
\label{app:proof:subseq}
\end{proof}

\section{Pruning for Temporal Subgraph Tests}
\label{app:pruning-temporal-test}

In order to further improve the speed, 
we adapt existing pruning techniques for subsequence matching to temporal subgraph tests.  

Injective mapping enumeration is the potential bottleneck of this algorithm.
As injective mapping enumeration is performed between node and enhanced node sequences without considering edge information,
false mappings could be enumerated, and the number of false mappings directly affects the speed of temporal subgraph tests.
Therefore, if false mappings can be detected earlier,
we can prune the search on those false mappings, and improve the speed on node mapping enumeration.

To reduce the number of false node mappings, 
we consider the following techniques to detect false node mappings, and prune unpromising enumeration branches.

\begin{itemize}
\item \stitle{Label sequence test}. Label sequence test detects false node mappings without considering node IDs.
After we replace the nodes in node, edge, and enhanced node sequences with node labels,
the subsequence relation between node and enhanced node sequences and the subsequence relation between edge sequences should hold,
if any temporal subgraph relation exists. 
Otherwise, we can safely claim no temporal subgraph relations exist.

\item \stitle{Local information match}. Local information match considers local structure information to decide node mapping instead of node label only.
The considered local structure information includes in/out node degree and in/out edge sequence.

\item \stitle{Prefix pruning}. Any partial node mapping forms a prefix. 
If a prefix has been searched before,
we can safely prune the search if we meet the prefix again. 
\end{itemize}

\section{Proof of Lemma 6}
\begin{proof}
We prove the sufficiency and necessity as follows.

First, we prove if $g_1 \subseteq_t g_2$ and $I(\mathbb{G}, g_1) = I(\mathbb{G}, g_2)$, then $\mathbb{R}(\mathbb{G}, g_1) = \mathbb{R}(\mathbb{G}, g_2)$.
Let $G'_2$ be the match of $g_2$ in $G$.
Since $g_1 \subseteq_t g_2$, there exists at least one subgraph $G'_1$ in $G'_2$ that matches $g_1$, where $|R(G, G'_1)| \geq |R(G, G'_2)|$.
Moreover, the subrelation $g_1 \subseteq_t g_2$ also implies $|\mathbb{R}(\mathbb{G}, g_1)| \geq |\mathbb{R}(\mathbb{G}, g_2)|$.
If $I(\mathbb{G}, g_1) = I(\mathbb{G}, g_2)$,  
then $|\mathbb{R}(\mathbb{G}, g_1)| = |\mathbb{R}(\mathbb{G}, g_2)|$,
and there exists a one-one mapping from the residual graphs in $\mathbb{R}(\mathbb{G}, g_1)$ to the residual graphs in  $\mathbb{R}(\mathbb{G}, g_2)$.
$\forall R(G, G'_1) \in \mathbb{R}(\mathbb{G}, g_1)$, its mapped $R(G, G'_2) \in \mathbb{R}(\mathbb{G}, g_2)$ has to satisfy $|R(G, G'_1)| = |R(G, G'_2)|$ such that $I(\mathbb{G}, g_1) = I(\mathbb{G}, g_2)$.
Therefore, $R(G, G'_1)$ is equivalent to $R(G, G'_2)$, and moreover, $\mathbb{R}(\mathbb{G}, g_1) = \mathbb{R}(\mathbb{G}, g_2)$.

Second, we prove if $g_1 \subseteq_t g_2$ and $\mathbb{R}(\mathbb{G}, g_1) = \mathbb{R}(\mathbb{G}, g_2)$, then $I(\mathbb{G}, g_1) = I(\mathbb{G}, g_2)$.
Since $\mathbb{R}(\mathbb{G}, g_1) = \mathbb{R}(\mathbb{G}, g_2)$, $g_1$ and $g_2$ share the same set of residual graphs.
Therefore, we can directly infer $I(\mathbb{G}, g_1) = I(\mathbb{G}, g_2)$.

In sum, we have proved Lemma~\ref{lm:residual-equivalence}.
\end{proof}

\section{Dataset Description}
\label{app:data-collection}

\stitle{Training data}.
Training data is a set of temporal graphs representing a set of syscall logs generated from a closed environment (\ie a server used to collect syscall logs with as little noise as possible), where security-related behaviors are performed.

First, we target at $12$ behaviors as representatives for the basic security-related behaviors that have drawn attention in cybersecurity study~\cite{sshattack, bayer2009view, invernizzi2014nazca, yin2007panorama}.
We execute the popular applications that perform these behaviors 
and collect the syscall logs as training data.
In general, we consider security-related behaviors from five categories.
\begin{enumerate*}
\item \emph{File compression/decompression} includes behaviors that compressed/decompress files using build-in compression software. 
These behaviors usually happen when malware compresses private information files for stealing information, 
or decompresses source code files to build any malicious binaries.
The cybersecurity application in Section~\ref{sec:intro} shows one example scenario where 
the graph query includes the behavior of file compression.
In this study, bzip2-based and gzip-based decompression behaviors are included.

\item \emph{Source code compilation} includes behaviors where source code files are complied into excitable binaries.
These behaviors are usually used by malware to build malicious binaries.
In this study, we focus on the behaviors of gcc-based and g++-based compilations.

\item \emph{File download/upload} are behaviors where system build-in software downloads files from remote servers, or uploads local files to remote servers.
When malware wants to start new types of attacks, it first needs to conduct file download to obtain new source code files for building new malicious binaries. 
Meanwhile, malware usually steals private information by uploading private files to unknown remote servers.
The behavior ``send-to-remote-server'' in the cybersecurity application in Section~\ref{sec:intro} belongs to such category.
In this study, the behaviors of ftp-based, scp-based, and wget-based download are considered.

\item \emph{Remote login} involves behaviors where a user logs into a remote server. 
For attackers, these behaviors are usually the first step for breaking into a computer system. 
The behavior ``sshd-login'' in the cybersecurity application in Section~\ref{sec:intro} falls into this category.
We include the behaviors of ssh-based login (client side), sshd-based login (server side), and ftpd-based login (server side) in this study.

\item \emph{System software management} includes behaviors where software applications are installed or updated by system build-in software.
These behaviors are conducted by attackers when they want to install malware into a computer system.
We investigate the behaviors of apt-get based update and apt-get based installation in this study.
\end{enumerate*}
For each target behavior, it is independently performed $100$ times in a closed environment with the guarantee that no other target behaviors are performed,
and accordingly, $100$ temporal graphs ($100$ syscall logs) are collected.
More details about the syscall logs for each behavior are shown in Table~\ref{table:data},
where the average number of nodes/edges in each syscall log is reported.

Second, we collected background syscall logs from the same closed environment.
In this process, the closed environment runs only the default applications, and none of the target behaviors are performed.
First, we kept the closed environment running for one week and collected the syscall logs.
Second, we randomly sampled $10,000$ temporal graphs from the syscall logs of the closed environment.
Each sampled temporal graph is based on a fixed time interval,
which is the longest time duration observed among the 12 target behaviors.
These $10,000$ temporal graphs are used as the background data.
The details of the background graph data are shown in Table~\ref{table:data},
where the averages of nodes/edges in the syscall logs are reported.

\stitle{Test data}.
Test data includes a large temporal graph representing a syscall log collected from another independent process.
In an ordinary desktop, in addition to the default workload the desktop hosts,
the following procedure is executed every one minute:
(1) one of the target behaviors is randomly selected and performed;
and (2) we record the time interval during which the selected behavior is performed as the ground truth.
In this way, we collected the syscall logs of $7$ days, 
which is a large temporal graph with $546,882$ nodes, $6,571,180$ edges.
This temporal graph contains $10,000$ behavior instances for the 12 target behaviors.
Based on the patterns discovered by our algorithms,
we formulate graph pattern queries to search those behavior instances from the test data.

Note that the closed environments for training data collection and test data collection are implemented in servers installed with Ubuntu 14.04.
But the techniques discussed in this paper are also applicable to other operating systems. 

\section{Domain knowledge based ranking function}
\label{app:domain}

When \tgmine returns multiple discriminative temporal graph patterns that have the same highest discriminative score,
the returned patterns are further ranked using the interest score of the nodes in the patterns.
The interest score of a node is defined using domain knowledge, 
and the interest score for a pattern is the sum of the interest scores for all the nodes in the pattern.
From all the returned temporal graph patterns,
the top-$5$ temporal graph patterns are used to formulate graph pattern queries for the target behavior. 

In this work,
we define the interest score as follows.
Given a node label $l$
\[
\interest(l) = \frac{1}{\freq(l)},
\]
where $\freq(l)$ is the number of temporal graphs in training data that contain this node label $l$.
Moreover,
we keep a blacklist.
For the node labels in the blacklist, such as ``TmpFile'', ``CacheFile'', ``/proc/stat/$*$'', and many other node labels that provide little security-related information,
we manually set their interest score as $0$.

\section{Scalability Evaluation}
\label{app:scalability}

\begin{figure}[tbh]
\center
\includegraphics[scale=0.7]{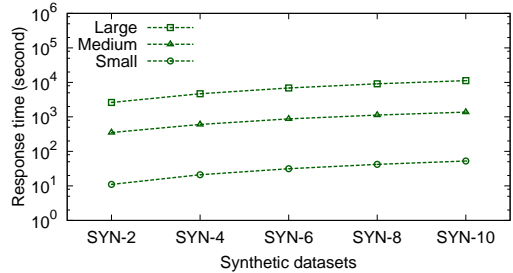}
\caption{Response time over synthetic datasets}
\label{fig:exp:efficiency:scale}
\end{figure}

Figure~\ref{fig:exp:efficiency:scale} shows the response time of \tgmine using synthetic datasets {\syn}-2, {\syn}-4, {\syn}-6, {\syn}-8, and {\syn}-10 as training data, respectively. The response time of \tgmine linearly scales with the size of training data. From the training data with up to 20M nodes and 80M edges, \tgmine can mine all discriminative patterns of up to 45 edges within 3 hours. 

%===========================Commented====================================
%===========================Commented====================================
%===========================Commented====================================
%===========================Commented====================================

\end{document}